\documentclass[aps,prl,twocolumn,showpacs,psfig,superscriptaddress,longbibliography]{revtex4-2}
\usepackage{amsfonts}
\usepackage{mathrsfs}
\usepackage{siunitx}
\usepackage{amsmath}% needed for subequations
\usepackage{color}
\usepackage{natbib}
\usepackage{textcomp}
\usepackage{graphicx}
\usepackage{bm}% bold maths
\usepackage{amssymb}
\usepackage{xspace}
\usepackage{epstopdf}
\usepackage{dcolumn}% Align table columns on decimal point
\usepackage{longtable}
\usepackage{setspace}
\usepackage{gensymb}
\usepackage{multirow}
\usepackage[colorlinks=true, letterpaper=true, pdfstartview=FitV, linkcolor=blue, citecolor=blue, urlcolor=blue]{hyperref}
\usepackage{float}
\usepackage{bm}

\usepackage{bibentry}

\makeatletter
\newcommand{\Rmnum}[1]{\expandafter\@slowromancap\romannumeral #1@}
\makeatother

\begin{document}

\title{Structural Alter-Phononics: Sublattice-Momentum Locking in Spinless Lattice Dynamics}

\author{Jing-Yang You}
\email{phyjyy@buaa.edu.cn}
\affiliation{Peng Huanwu Collaborative Center for Research and Education, Beihang University, Beijing 100191, China}

\author{Zhen Zhang}
\affiliation{State Key Laboratory of Quantum Optics and Quantum Optics Devices, Institute of Theoretical Physics, Shanxi University, Taiyuan 030006, China}

\author{Xianlei Sheng}
\email{xlsheng@buaa.edu.cn}
\affiliation{School of Physics, Beihang University, Beijing 100191, China}
\affiliation{Peng Huanwu Collaborative Center for Research and Education, Beihang University, Beijing 100191, China}

\author{Gang Su}
\email{sugang@itp.ac.cn}
\affiliation{Institute of Theoretical Physics, Chinese Academy of Sciences, Beijing 100190, China}

\begin{abstract}
The discovery of altermagnetism has shown that crystal symmetry can generate momentum-dependent internal polarization without net magnetization. Whether an analogous form of symmetry-organized momentum-space order can exist for spinless lattice vibrations remains unresolved. Here we identify a structural mechanism for \emph{alter-phononics}, in which phonon eigenmodes formed from structurally equivalent sublattices acquire momentum-dependent sublattice polarization and frequency splitting in nonmagnetic crystals. The central quantity is the sublattice-resolved dynamical asymmetry $\Delta(\mathbf q)=D_{AA}(\mathbf q)-D_{BB}(\mathbf q)$, which controls the associated eigenvector polarization. We show that this effect requires an alter-generator that maps equivalent sublattices onto one another while rotating the wave vector, together with the absence of inversion exchange and little-group sublattice-exchange constraints that would otherwise enforce sublattice equipartition. These symmetry rules generate nematic $d$-wave, tetragonal $d/g$-wave, and tripartite six-lobe phonon textures. First-principles calculations demonstrate the mechanism in representative nonmagnetic crystals and show how a symmetry-preserving structural distortion can unlock a hidden $d_{x^2-y^2}$-type texture by removing glide-induced equipartition traps while retaining the screw-axis alter-generator. We further show that the eigenvector texture is inherited by sublattice-projected electron-phonon coupling and anharmonic response functions. Our results establish structural alter-phononics as a spinless counterpart to altermagnetic momentum-space order and provide experimentally testable signatures in finite-$\mathbf q$ phonon spectra and displacement patterns.
\end{abstract}
\maketitle

{\color{blue}Introduction} -- Symmetry-enforced momentum-space polarization is a central theme in modern condensed matter physics. In electronic systems, inversion-symmetry breaking produces Rashba- or Dresselhaus-type spin textures through relativistic spin--orbit coupling~\cite{Rashba,Dresselhaus1955,LaShell1996,Manchon2015}, whereas time-reversal-symmetry breaking produces magnetic exchange splitting~\cite{Kittel}. Altermagnetism has recently revealed a different route to momentum-space spin polarization: compensated magnetic sublattices can generate large momentum-dependent spin splitting without net magnetization, because crystal operations connect opposite-spin sublattices while rotating the wave vector~\cite{Smejkal2022,Smejkal2022a,Mazin2022,Bai2024,Zhang2024,Guo2025,Huang2026,Zhu2026}. The resulting spin splitting is not an accidental consequence of low symmetry, but a symmetry-organized $d$-, $g$-, or $i$-wave texture in reciprocal space~\cite{Hayami2019,Ahn2019,GonzalezHernandez2021,Leeb2024,Krempasky2024}. Experimental and materials studies have identified altermagnetic behavior in systems such as MnTe, RuO$_2$, and CrSb~\cite{Krempasky2024,Lee2024,Reimers2024}, as well as in more complex compounds such as KV$_2$Se$_2$O~\cite{Jiang2025}. Related symmetry-organized momentum-space structures have also been discussed for other quasiparticle and collective excitations~\cite{Cao2025}.

Phonons provide a conceptually different arena for exploring symmetry-organized momentum-space textures. They are spinless bosonic excitations governed by a dynamical matrix rather than by an electronic Hamiltonian with exchange fields. Nevertheless, phonon eigenvectors carry rich internal information, including atomic displacement patterns, sublattice weights, circular polarization, and angular momentum. Topological nodal-line and Weyl phonons~\cite{Zhang2018,Jin2018,Liu2019}, symmetry-based phonon topology~\cite{Xu2024}, and chiral phonons carrying angular momentum~\cite{Zhang2014,Zhu2018,Ishito2022} have established that lattice dynamics are highly sensitive to crystalline symmetry. The symmetry conditions for phonon angular momentum have also been systematically classified~\cite{Coh2023}, and phonon angular momentum can couple to magnetic degrees of freedom and external fields, giving rise to phonon magnetic moments, phonon Zeeman effects, and phonon thermal Hall responses~\cite{Strohm2005,Sheng2006,Mori2014,Li2020,Grissonnanche2020,Chen2020}. Recent studies have further connected altermagnetic order with lattice dynamics through dynamic paramagnon-polarons and magnon--phonon-induced phonon angular-momentum textures~\cite{Steward2023,Bendin2025}. More generally, alteraxial phonons in collinear magnets illustrate how magnetic point-group symmetry can organize higher-order momentum-space textures of phonon angular momentum~\cite{WangAlteraxial2026}. These developments motivate a broader question of whether analogous symmetry organization can arise directly in the spinless lattice-dynamical degrees of freedom of nonmagnetic crystals.

\begin{figure}[!!htbp]
  \centering
  \includegraphics[scale=0.32,angle=0]{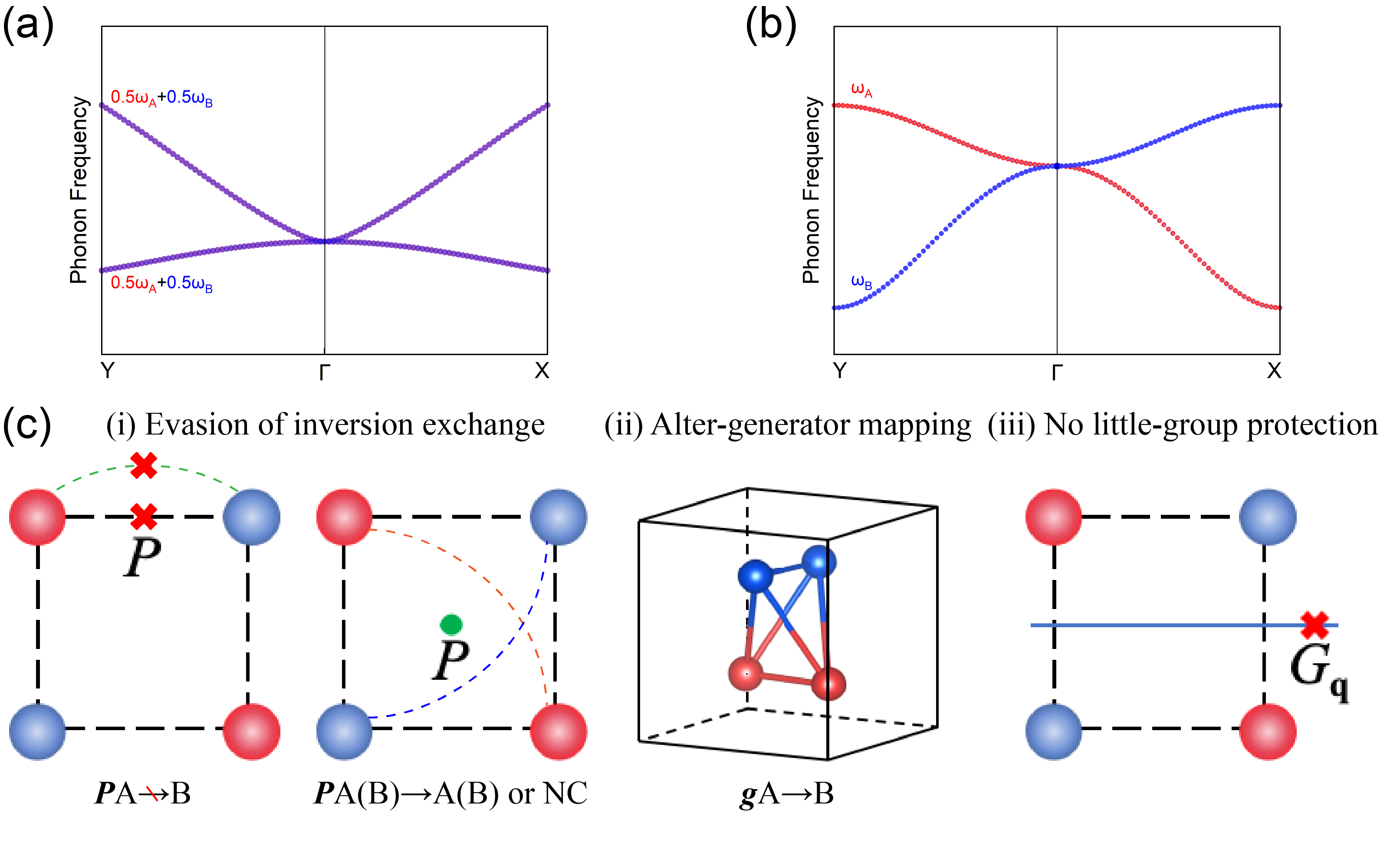}\\
  \caption{\textbf{Concept and symmetry criteria of structural alter-phononics.}
(a) Conventional phonon modes formed by two structurally equivalent sublattices A and B. When symmetry enforces $D_{AA}(\mathbf q)=D_{BB}(\mathbf q)$, the eigenmodes exhibit sublattice equipartition, with equal A/B vibrational weights and vanishing sublattice polarization $P(\mathbf q)=0$.
(b) Alter-phononic modes. Once the local dynamical asymmetry $\Delta(\mathbf q)=D_{AA}(\mathbf q)-D_{BB}(\mathbf q)$ is symmetry-allowed, the two branches acquire momentum-dependent frequency splitting and opposite sublattice polarization. The dominant sublattice character switches between symmetry-related momentum directions, forming sublattice-momentum locking.
(c) Symmetry requirements for structural alter-phononics. First, global inversion exchange between A and B must be absent; otherwise spinless time-reversal symmetry enforces sublattice equipartition throughout the Brillouin zone. Second, an alter-generator $g=\{R|\boldsymbol\tau\}$ must connect A and B while rotating the wave vector, imposing $D_{AA}(\mathbf q)=D_{BB}(R\mathbf q)$. Third, little-group operations that locally exchange A and B along the target momentum path must be absent; otherwise they create sublattice-equipartition, or polarization-nodal, lines.}\label{fig1}
\end{figure}

Here we show that a nonmagnetic phonon mode can exhibit an altermagnetic-like momentum-space texture through its sublattice-resolved displacement character. The relevant internal degree of freedom is neither electronic spin nor phonon angular momentum, but the sublattice character of the phonon eigenvector. In a conventional phonon mode formed from two structurally equivalent sublattices, symmetry often enforces equal vibrational weights on the two sublattices [Fig.~\ref{fig1}(a)]. In an alter-phononic state, by contrast, the two sublattices remain equivalent in real space but become dynamically inequivalent at a given momentum. A phonon branch can therefore be dominated by sublattice A along one momentum direction and by sublattice B along a symmetry-related direction [Fig.~\ref{fig1}(b)].

The key physical object is the sublattice-resolved dynamical asymmetry
$\Delta(\mathbf q)=D_{AA}(\mathbf q)-D_{BB}(\mathbf q)$,
which controls both the momentum-dependent frequency splitting and the sublattice polarization of the phonon eigenvectors. A finite $\Delta(\mathbf q)$ is symmetry-allowed only when local sublattice equipartition is not enforced at the target momentum. At the same time, a macroscopic alter-generator must remain: a space-group operation $g=\{R|\boldsymbol\tau\}$ that maps the active sublattices onto one another while rotating $\mathbf q$. This operation imposes the relation
$D_{AA}(\mathbf q)=D_{BB}(R\mathbf q)$,
so that the sublattice identity of the phonon branch switches between symmetry-related momentum directions. This sublattice--momentum locking is the defining mechanism of structural alter-phononics.

This mechanism provides a symmetry-based route to phonon branch polarization beyond ordinary anisotropic phonon dispersion. A low-symmetry environment may lift phonon degeneracies, but structural alter-phononics requires a more specific sublattice constraint: the active sublattices must not be locally equipartitioned by inversion or by little-group operations along the target momentum path, while an alter-generator must still connect them globally by rotating the wave vector. These requirements make alter-phononics a combined space-group-and-Wyckoff phenomenon rather than a space-group property alone.

We formulate this mechanism using a minimal dynamical-matrix theory and derive the symmetry rules governing the allowed textures. Because nonmagnetic phonons obey time-reversal symmetry, the resulting frequency splitting is even in $\mathbf q$, leading to nematic $d$-wave, tetragonal $d/g$-wave, and tripartite six-lobe textures rather than odd-parity two-sublattice splittings. First-principles calculations demonstrate these textures in representative nonmagnetic crystals. A tetragonal prototype further shows how a structural distortion can unlock a hidden $d_{x^2-y^2}$-type texture by breaking glide-induced equipartition traps while preserving the screw-axis alter-generator. Additional examples demonstrate the hierarchy of alter-generator-imposed textures. Finally, we show that the same eigenvector texture enters sublattice-projected electron--phonon coupling and anharmonic response functions, providing finite-$\mathbf q$ experimental signatures beyond phonon band dispersions.

{\color{blue}Dynamical-matrix mechanism and symmetry constraints} -- We begin with a minimal description of two local vibrational modes associated with structurally equivalent sublattices A and B. Projecting the harmonic force-constant problem into this two-mode subspace gives the mass-weighted dynamical matrix
\begin{equation}
D(\mathbf q)=
\begin{pmatrix}
D_{AA}(\mathbf q) & D_{AB}(\mathbf q)\\
D_{BA}(\mathbf q) & D_{BB}(\mathbf q)
\end{pmatrix},
\label{eq:Dq}
\end{equation}
where $D_{AA}$ and $D_{BB}$ describe the effective intra-sublattice restoring forces and $D_{AB}=D_{BA}^{*}$ describes inter-sublattice hybridization. The eigenfrequencies are
\begin{equation}
\omega_{\pm}^2(\mathbf q)=
\frac{D_{AA}(\mathbf q)+D_{BB}(\mathbf q)}{2}
\pm
\sqrt{
\left[\frac{\Delta(\mathbf q)}{2}\right]^2+|D_{AB}(\mathbf q)|^2
},
\label{eq:omega_pm}
\end{equation}
with $\Delta(\mathbf q)=D_{AA}(\mathbf q)-D_{BB}(\mathbf q)$. The quantity $\Delta(\mathbf q)$ measures whether two sublattices that are equivalent in real space become dynamically inequivalent at momentum $\mathbf q$. It is therefore the central order parameter of structural alter-phononics.

The corresponding sublattice polarization of the two branches can be written as
\begin{equation}
P_{\pm}(\mathbf q)=
\frac{W_A^{\pm}(\mathbf q)-W_B^{\pm}(\mathbf q)}
{W_A^{\pm}(\mathbf q)+W_B^{\pm}(\mathbf q)}
=
\pm\frac{\Delta(\mathbf q)}{
\sqrt{\Delta^2(\mathbf q)+4|D_{AB}(\mathbf q)|^2}}
\label{eq:polarization}
\end{equation}
in the minimal two-mode model. This equation makes explicit that frequency splitting and sublattice polarization are related but not identical. A finite splitting can also arise from $D_{AB}$, whereas a finite $P_{\pm}$ requires $\Delta(\mathbf q)\neq0$. Thus, the defining fingerprint of alter-phononics is not a phonon splitting alone, but the symmetry-organized switching of sublattice character in momentum space, as indicated schematically in Fig.~\ref{fig1}(b).

Three symmetry constraints determine whether $\Delta(\mathbf q)$ can be nonzero and whether it forms a macroscopic texture. These constraints are summarized in Fig.~\ref{fig1}(c).

First, global inversion exchange forbids the effect. If inversion maps A to B, then $D_{AA}(\mathbf q)=D_{BB}(-\mathbf q)$. For spinless nonmagnetic phonons, time reversal gives $D_{BB}(-\mathbf q)=D_{BB}(\mathbf q)$, and therefore $\Delta(\mathbf q)=0$ throughout the Brillouin zone. Centrosymmetric candidate materials can still host structural alter-phonons only when the active sublattices are not exchanged by inversion.

Second, an alter-generator must connect the active sublattices while rotating the wave vector. For a space-group operation $g=\{R|\boldsymbol\tau\}$ satisfying $gA=B$, symmetry imposes $D_{AA}(\mathbf q)=D_{BB}(R\mathbf q)$. When $R\mathbf q$ is not locally equivalent to $\mathbf q$, this relation does not force $D_{AA}(\mathbf q)=D_{BB}(\mathbf q)$. Instead, it organizes the sublattice asymmetry so that a branch dominated by A at $\mathbf q$ is related to a branch dominated by B at $R\mathbf q$. This is the phononic analogue of altermagnetic momentum-space polarization, with sublattice weight replacing spin.

Third, local little-group operations can quench the effect along selected paths. If an operation $h\in G_{\mathbf q}$ leaves $\mathbf q$ invariant while exchanging A and B, it enforces $D_{AA}(\mathbf q)=D_{BB}(\mathbf q)$ and hence $P_{\pm}(\mathbf q)=0$. Such paths become polarization nodal lines. They need not be frequency-degeneracy lines, because the splitting may remain finite through $D_{AB}(\mathbf q)$. A true frequency degeneracy requires additional symmetry constraints that also suppress the relevant hybridization.

These rules show that structural alter-phononics is not merely a consequence of low symmetry. The alter-generator creates the macroscopic texture, while inversion exchange and little-group exchange determine where the texture is forbidden. As a result, compatible space groups alone are insufficient: the active atoms must occupy Wyckoff positions that avoid the sublattice-exchange operations that restore equipartition. Table~\ref{tab1} summarizes representative space groups, alter-generators, and Wyckoff constraints for nematic $d$-wave, tetragonal $d/g$-wave, tripartite six-lobe, and $i$-wave alter-phononic textures.

\begin{table*}[!hbtp]
\footnotesize
\centering
\caption{\textbf{Symmetry design principles for structural alter-phononics.} For each wave symmetry, representative space groups (SGs) are categorized into non-centrosymmetric (top sub-row) and centrosymmetric (bottom sub-row) classes, demonstrating the parity-independent nature of alter-phonons. To avoid trivial sublattice-equipartition traps, atoms must satisfy stringent Wyckoff constraints. \textit{Evasion of global inversion}: in centrosymmetric SGs, equivalent sublattices must not be interchanged by global spatial inversion $\mathcal{P}$. \textit{Off-axis/off-plane occupation}: atoms must deviate from symmetry-enforcing $C_n$ axes, mirror, or glide planes. \textit{Off-principal-axis occupation}: in tripartite and $i$-wave systems, atoms must possess finite displacements from the primary rotational axes to preserve distinct, symmetry-linked sublattices within the primitive cell. Enantiomorphic operations (e.g., $4_1$ and $4_3$) yielding identical macroscopic topologies are grouped together.}
\label{tab1}
\begin{tabular}{c c c c c}
\hline
Symmetry & Crystal system & Alter-generator(s) & Representative SGs (Top: NC / Bottom: C) & Wyckoff constraints \\
\hline
Nematic $d$-wave & Monoclinic/ & $C_2,\; 2_1,\; m,\; n$ & 4 ($P2_1$), 18 ($P2_12_12$), 33 ($Pna2_1$) & Off-axis/off-plane\\
(Quasi-1D, 4-leaf) & Orthorhombic & & 10 ($P2/m$), 14 ($P2_1/c$), 58 ($Pnnm$) & occupation \\
\hline
$d$-wave ($l=2$), & Tetragonal & $C_4,\; 4_1\,(4_2, 4_3),\; \bar{4}$ & 75 ($P4$), 81 ($P\bar{4}$), 111 ($P\bar{4}2m$), 121 ($I\bar{4}2d$) & Evasion of global\\
$g$-wave ($l=4$) & & & 83 ($P4/m$), 84 ($P4_2/m$), 87 ($I4/m$), 136 ($P4_2/mnm$) & inversion \\
\hline
Tripartite 6-leaf & Trigonal/ & $C_3,\; 3_1\,(3_2),\; \bar{3}$ & 143 ($P3$), 149 ($P321$), 156 ($P3m1$), 198 ($P2_13$) & Off-principal-axis\\
(Even parity) & Cubic & & 147 ($P\bar{3}$), 167 ($R\bar{3}c$), 205 ($Pa\bar{3}$) & occupation \\
\hline
$i$-wave ($l=6$) & Hexagonal & $C_6,\; 6_1\,(6_3, 6_5),\; \bar{6}$ & 168 ($P6$), 174 ($P\bar{6}$), 186 ($P6_3mc$) & Off-principal-axis\\
(12-leaf) & & & 175 ($P6/m$), 176 ($P6_3/m$), 194 ($P6_3/mmc$) & occupation \\
\hline
\end{tabular}
\end{table*}

\begin{figure}[!!htbp]
  \centering
  \includegraphics[scale=0.102,angle=0]{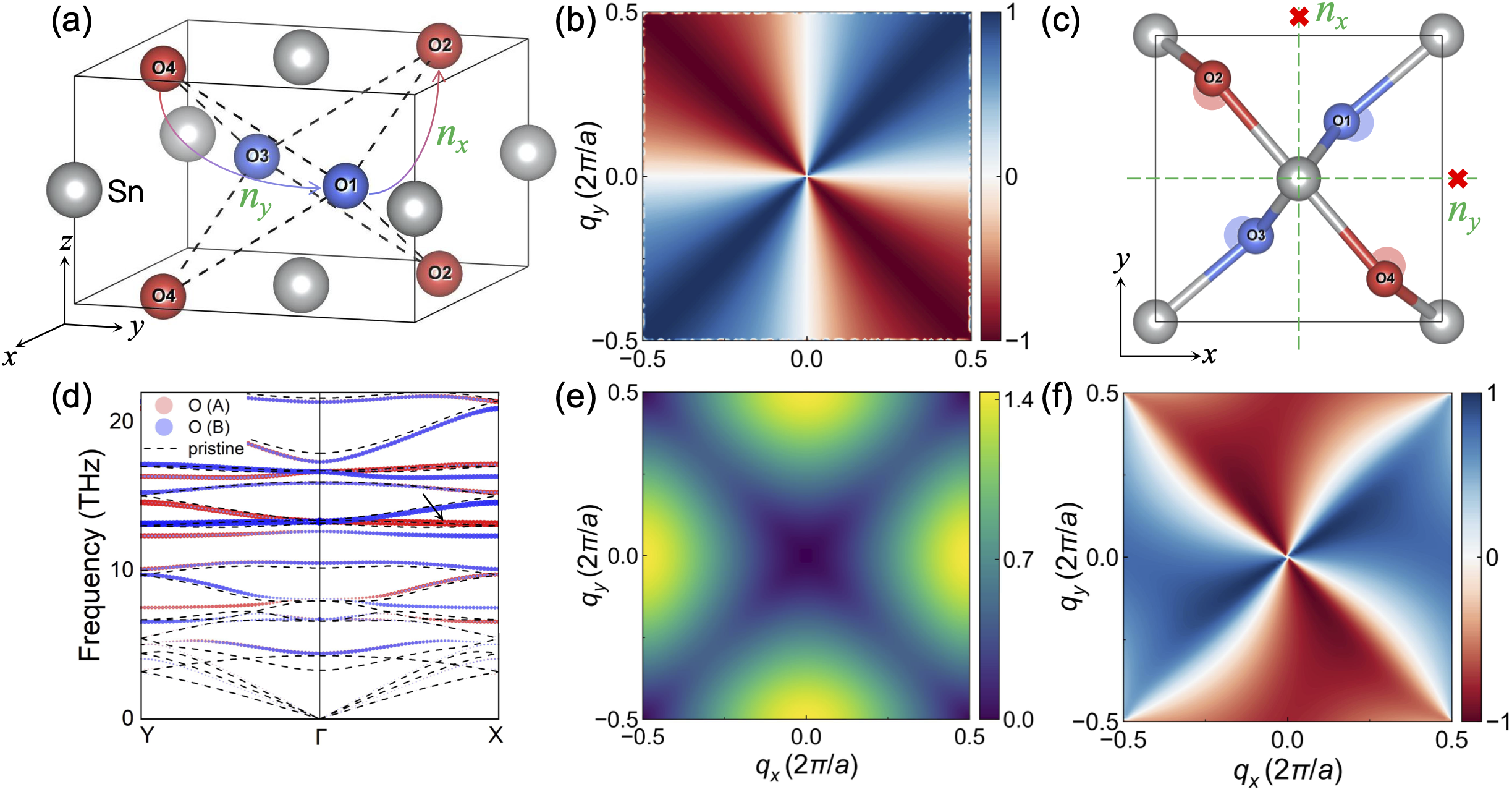}\\
  \caption{\textbf{Symmetry unlocking of a tetragonal $d$-wave alter-phonon texture.}
(a) Pristine rutile SnO$_2$ in space group $P4_2/mnm$ (SG 136). The two active oxygen sublattices are related by a $4_2$ screw alter-generator, but the $n_x/n_y$ glide planes locally exchange the sublattices along the principal momentum axes.
(b) Sublattice-polarization map of the pristine phase, $P(\mathbf q)=[W_A(\mathbf q)-W_B(\mathbf q)]/[W_A(\mathbf q)+W_B(\mathbf q)]$. The glide symmetries force $P(\mathbf q)=0$ along $\Gamma$--X and $\Gamma$--Y, so the sublattice-polarized antinodes appear along the diagonal directions, giving a hidden $d_{xy}$-type texture.
(c) Distorted structure in $P4_2/m$ (SG 84), obtained by an asymmetric in-plane torsion of the SnO$_6$ octahedra. The distortion breaks the glide-induced equipartition constraints while preserving the $4_2$ screw alter-generator.
(d) Phonon dispersions of the pristine and distorted phases, highlighting the unlocked splitting along the principal axes.
(e) Momentum-resolved frequency splitting $\Delta\omega(\mathbf q)$ of the selected optical branches in the distorted phase.
(f) Corresponding sublattice-polarization map, showing the unlocked $d_{x^2-y^2}$ texture. The same phonon branch is dominated by sublattice A along one principal direction and by sublattice B along the orthogonal direction.}\label{fig2}
\end{figure}

{\color{blue}Symmetry unlocking of a tetragonal $d$-wave texture} -- A tetragonal prototype illustrates how structural alter-phononics can be hidden by little-group constraints and unlocked by a targeted distortion. In rutile SnO$_2$, the high-symmetry structure belongs to $P4_2/mnm$ (SG 136). The active oxygen sites form two equivalent sublattices connected by a $4_2$ screw operation, as shown in Fig.~\ref{fig2}(a). This screw operation is the relevant alter-generator: it maps A to B while rotating the in-plane momentum, imposing $D_{AA}(q_x,q_y)=D_{BB}(-q_y,q_x)$. If this were the only constraint, it would generate a tetragonal sublattice texture.

However, the pristine structure also contains $n_x$ and $n_y$ glide symmetries. Along $\Gamma$--X, where $q_y=0$, the $n_y$ glide leaves the path invariant while exchanging the two oxygen sublattices. It therefore enforces
\begin{equation}
D_{AA}(q_x,0)=D_{BB}(q_x,0),\qquad \Delta(q_x,0)=0.
\end{equation}
The same holds along $\Gamma$--Y. Thus the principal axes are sublattice-equipartition lines. The alter-phononic tendency associated with the $4_2$ screw is not absent, but it is hidden: the sublattice-polarized antinodes appear along the diagonal directions, producing a $d_{xy}$-type texture, as shown by the polarization map in Fig.~\ref{fig2}(b).

We then introduce an asymmetric in-plane torsion of the SnO$_6$ octahedra, lowering the symmetry from SG 136 to $P4_2/m$ (SG 84), as shown in Fig.~\ref{fig2}(c). This distortion breaks the glide-induced equipartition constraints but preserves the $4_2$ screw alter-generator. Consequently, $\Delta(q_x,0)$ and $\Delta(0,q_y)$ become symmetry-allowed, while the screw still relates the two principal directions:
\begin{equation}
D_{AA}(q_x,0)=D_{BB}(0,q_x),\qquad
D_{BB}(q_x,0)=D_{AA}(0,q_x).
\end{equation}
The result is an unlocked $d_{x^2-y^2}$-type alter-phonon texture: a selected optical branch is dominated by oxygen sublattice A along one principal direction and by sublattice B along the orthogonal direction. First-principles phonon calculations and fat-band projections confirm this evolution in Figs.~\ref{fig2}(d)--\ref{fig2}(f). This example isolates the essential design rule: remove the symmetries that enforce local equipartition, but preserve the alter-generator that dictates the momentum-space switching of sublattice identity.

\begin{figure}[!!htbp]
  \centering
  \includegraphics[scale=0.25,angle=0]{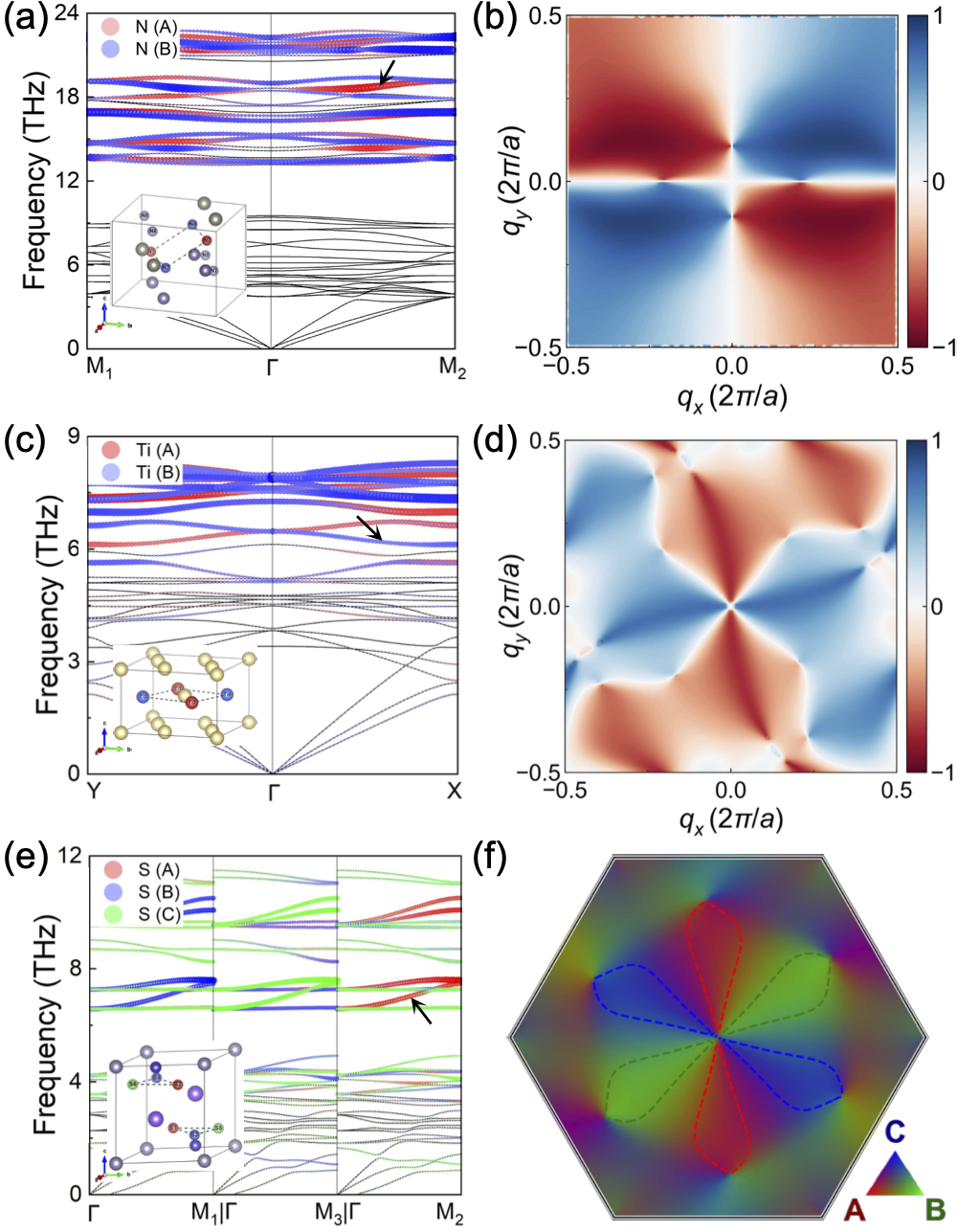}\\
  \caption{\textbf{Hierarchy of alter-generator-imposed phonon textures.}
Sublattice-resolved phonon dispersions and momentum-space polarization maps for representative nonmagnetic crystals.
(a,b) Orthorhombic ZnGeN$_2$ (SG 33), where a twofold sublattice-mapping operation produces a nematic $d$-wave texture. The reduced rotational symmetry leads to an anisotropic four-lobe pattern.
(c,d) Tetragonal Ti$_2$Ga$_3$ (SG 83), where a tetragonal alter-generator produces a conventional $d_{x^2-y^2}$-type texture. The dominant sublattice character reverses between orthogonal momentum directions.
(e,f) Trigonal K$_2$SnAs$_2$S$_6$ (SG 147), where three symmetry-related active sublattices generate a tripartite six-lobe texture. The ternary color map represents the fractional weights $W_i/\sum_j W_j$ with $i=A,B,C$. The identical color pattern at $\mathbf q$ and $-\mathbf q$ confirms compatibility with spinless time-reversal symmetry.
Black arrows mark the optical branches selected for the two-dimensional momentum-space maps.}\label{fig3}
\end{figure}

{\color{blue}Hierarchy of alter-generator-imposed textures} -- The SnO$_2$ prototype demonstrates symmetry unlocking in a single tetragonal setting. We next show that the same principle organizes a hierarchy of structural alter-phonon textures across different crystal systems. Figure~\ref{fig3} compares orthorhombic ZnGeN$_2$, tetragonal Ti$_2$Ga$_3$, and trigonal K$_2$SnAs$_2$S$_6$, which realize, respectively, a nematic $d$-wave texture, a conventional tetragonal $d$-wave texture, and a tripartite six-lobe texture.

In ZnGeN$_2$, the relevant sublattice-mapping operation has an effective twofold character. The reduced rotational symmetry produces a strongly anisotropic, nematic $d$-wave sublattice texture rather than an isotropic four-lobe pattern, as shown in Fig.~\ref{fig3}(a,b). In Ti$_2$Ga$_3$, a tetragonal alter-generator organizes the eigenvector weights into a conventional $d_{x^2-y^2}$-type pattern: the dominant sublattice character reverses between orthogonal momentum directions, as shown in Fig.~\ref{fig3}(c,d). In K$_2$SnAs$_2$S$_6$, three symmetry-related active sublattices generate a tripartite texture described by the fractional weight vector $\left\{\frac{W_A}{\sum_i W_i},\frac{W_B}{\sum_i W_i},\frac{W_C}{\sum_i W_i}\right\}$. The resulting six-lobe map in Fig.~\ref{fig3}(e,f) is even under $\mathbf q\to-\mathbf q$, as required by time reversal for spinless phonons.

These examples show that the rotational character of the alter-generator determines the macroscopic texture: twofold operations produce nematic $d$-wave patterns, tetragonal operations produce conventional four-lobe textures, and trigonal operations produce tripartite six-lobe textures. In each case, the central observable is the correlated evolution of phonon frequency and sublattice-resolved eigenvector character. Frequency splitting identifies where branches separate energetically; sublattice polarization identifies whether the branches carry the symmetry-required internal texture.

\begin{figure}[!!htbp]
  \centering
  \includegraphics[scale=0.18,angle=0]{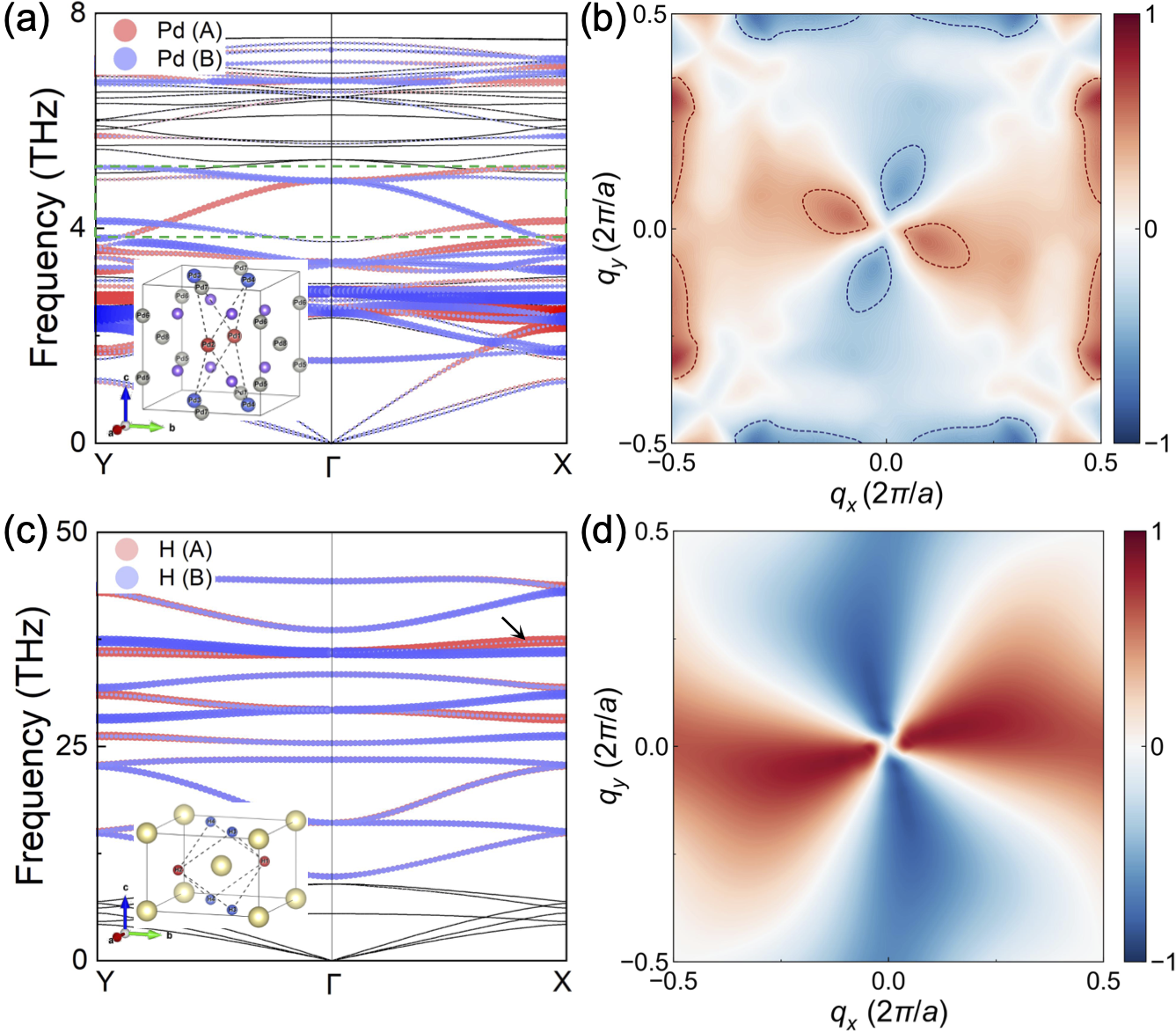}\\
  \caption{\textbf{Momentum-resolved response functions generated by alter-phonon textures.}
(a) Sublattice-resolved phonon dispersion of tetragonal PdSe (SG 84). The shaded frequency window marks the optical modes used to evaluate the sublattice-projected electron-phonon coupling.
(b) Momentum-space map of the coupling polarization $P_\lambda(\mathbf q)=[\lambda_A(\mathbf q)-\lambda_B(\mathbf q)]/[\lambda_A(\mathbf q)+\lambda_B(\mathbf q)]$. The $d$-wave pattern shows that electron-phonon scattering at symmetry-related momentum transfers proceeds through different sublattice vibration channels, although the total response preserves the crystal symmetry.
(c) Phonon dispersion of tetragonal MgH$_2$ (SG 84). The arrow marks the single hydrogen-dominated optical branch selected for the anharmonic analysis.
(d) Normalized anharmonic-polarization texture $P_\gamma(\mathbf q)=\widetilde{\gamma}(\mathbf q)P(\mathbf q)$, where $\widetilde{\gamma}(\mathbf q)=\gamma(\mathbf q)/\max_{\mathbf q}|\gamma(\mathbf q)|$ is the normalized mode Gr\"uneisen parameter of the selected branch and $P(\mathbf q)$ is its sublattice polarization. This definition preserves the Gr\"uneisen amplitude while encoding the sublattice identity of the same phonon eigenmode.}\label{fig4}
\end{figure}

{\color{blue}Momentum-resolved response functions} -- The sublattice texture of an alter-phonon is not only a property of the harmonic phonon eigenvector. Because phonon eigenvectors enter electron-phonon matrix elements, anharmonic scattering amplitudes, and dynamical structure factors~\cite{Giustino2017}, sublattice-momentum locking can be transferred to experimentally relevant response functions. Figure~\ref{fig4} demonstrates this transfer for sublattice-projected electron-phonon coupling in PdSe and sublattice-resolved anharmonicity in MgH$_2$.

For electron-phonon coupling, we define a sublattice-projected coupling strength $\lambda_i(\mathbf q)=\sum_{\nu\in\Omega}W_i^\nu(\mathbf q)\lambda_{\mathbf q\nu}$, where $W_i^\nu(\mathbf q)=\sum_{\kappa\in i,\alpha}|e_{\kappa\alpha}^\nu(\mathbf q)|^2$ is the eigenvector weight on sublattice $i$, and $\Omega$ denotes the selected phonon-frequency window. The normalized coupling polarization is $
P_\lambda(\mathbf q)=
\frac{\lambda_A(\mathbf q)-\lambda_B(\mathbf q)}
{\lambda_A(\mathbf q)+\lambda_B(\mathbf q)}$. First-principles calculations for tetragonal PdSe show a clear $d$-wave pattern in $P_\lambda$: scattering at one momentum direction is dominated by vibrations on sublattice A, whereas the symmetry-related direction is dominated by sublattice B, as shown in Fig.~\ref{fig4}(a,b). The total response still respects crystal symmetry, but its microscopic sublattice channel becomes momentum selective. This hidden structure becomes important when electronic states near the Fermi level also carry sublattice or orbital selectivity, enabling momentum-dependent quasiparticle renormalization and phonon linewidths~\cite{Gerber2017,Valla1999,Lanzara2001}.

For the anharmonic response in Fig.~\ref{fig4}(c,d), we focus on a single selected hydrogen-dominated optical branch rather than a frequency window containing multiple phonons. Therefore, we do not partition the mode Gr\"uneisen parameter by multiplying it with the sublattice weight $W_i(\mathbf q)$, which would artificially reduce the magnitude of the plotted quantity. Instead, we use the mode Gr\"uneisen parameter of the selected branch $\gamma(\mathbf q)=-\frac{\partial\ln\omega(\mathbf q)}{\partial\ln V}$, and normalize it over the plotted momentum plane $\widetilde{\gamma}(\mathbf q)=\frac{\gamma(\mathbf q)}{\max_{\mathbf q}|\gamma(\mathbf q)|}$.
The directional anharmonic texture is then characterized by $P_\gamma(\mathbf q)=\widetilde{\gamma}(\mathbf q) P(\mathbf q)$, where $P(\mathbf q)=[W_A(\mathbf q)-W_B(\mathbf q)]/[W_A(\mathbf q)+W_B(\mathbf q)]$ is the sublattice polarization of the same phonon branch. This construction separates the two pieces of information: $\widetilde{\gamma}(\mathbf q)$ measures the momentum dependence of the anharmonic frequency response, while $P(\mathbf q)$ encodes which sublattice dominates the corresponding eigenvector. Their product therefore visualizes the sublattice-resolved anisotropy of the anharmonic response without suppressing the Gr\"uneisen amplitude by an additional projection factor.

For tetragonal MgH$_2$, the selected hydrogen-dominated optical branch displays a $d_{x^2-y^2}$-type anharmonic texture, as shown in Fig.~\ref{fig4}(c,d). This indicates that acoustic phonons propagating along different directions can scatter through different sublattice components of the optical phonon bath. In equilibrium, the total thermal conductivity must obey the macroscopic crystal symmetry; however, under selective nonequilibrium driving of one sublattice component, for example using nonlinear phononics or local strain engineering~\cite{Foerst2011,Mankowsky2016}, the pre-existing sublattice-momentum locking can provide a route to directional control of phonon scattering~\cite{Li2012,Wehmeyer2017}.

{\color{blue}Discussion and conclusion} -- Structural alter-phononics should be distinguished from two related but different phenomena. It is not a generic low-symmetry phonon splitting, because its defining signature is the symmetry-locked switching of sublattice character between momentum directions connected by an alter-generator. It is also distinct from alteraxial phonons in collinear magnets~\cite{WangAlteraxial2026}: alteraxial phonons are classified by magnetic point groups and by momentum-space textures of phonon angular momentum, whereas structural alter-phonons are nonmagnetic, spinless, and characterized by sublattice-resolved dynamical asymmetry.

This distinction leads to different experimental signatures. The key observable is not only a finite-$\mathbf q$ phonon splitting, but the correlated switching of atomic displacement character between symmetry-related momenta. Momentum-resolved inelastic X-ray scattering, neutron scattering, and electron energy-loss spectroscopy are natural probes, because their cross sections depend on the dynamical structure factor and hence on the phonon displacement pattern~\cite{Miao2018,baron2020,Senga2019}. In practice, the combination of finite-$\mathbf q$ frequency splitting and symmetry-related switching of sublattice displacement character is the most direct fingerprint of structural alter-phononics.

In summary, we have identified structural alter-phononics as a spinless route to altermagnetic-like momentum-space order in lattice dynamics. The mechanism requires an alter-generator that maps equivalent sublattices onto one another while rotating the wave vector, together with the absence of symmetry operations that enforce sublattice equipartition. The resulting sublattice-momentum locking produces nematic $d$-wave, tetragonal $d/g$-wave, and tripartite six-lobe phonon textures in nonmagnetic crystals. First-principles calculations verify these textures and show that they can be inherited by electron-phonon and anharmonic response functions. These results provide a symmetry principle for designing phonon eigenvector textures and for controlling momentum-resolved lattice-mediated responses.


\begin{thebibliography}{50}%
\makeatletter
\providecommand \@ifxundefined [1]{%
 \@ifx{#1\undefined}
}%
\providecommand \@ifnum [1]{%
 \ifnum #1\expandafter \@firstoftwo
 \else \expandafter \@secondoftwo
 \fi
}%
\providecommand \@ifx [1]{%
 \ifx #1\expandafter \@firstoftwo
 \else \expandafter \@secondoftwo
 \fi
}%
\providecommand \natexlab [1]{#1}%
\providecommand \enquote  [1]{``#1''}%
\providecommand \bibnamefont  [1]{#1}%
\providecommand \bibfnamefont [1]{#1}%
\providecommand \citenamefont [1]{#1}%
\providecommand \href@noop [0]{\@secondoftwo}%
\providecommand \href [0]{\begingroup \@sanitize@url \@href}%
\providecommand \@href[1]{\@@startlink{#1}\@@href}%
\providecommand \@@href[1]{\endgroup#1\@@endlink}%
\providecommand \@sanitize@url [0]{\catcode `\\12\catcode `\$12\catcode
  `\&12\catcode `\#12\catcode `\^12\catcode `\_12\catcode `\%12\relax}%
\providecommand \@@startlink[1]{}%
\providecommand \@@endlink[0]{}%
\providecommand \url  [0]{\begingroup\@sanitize@url \@url }%
\providecommand \@url [1]{\endgroup\@href {#1}{\urlprefix }}%
\providecommand \urlprefix  [0]{URL }%
\providecommand \Eprint [0]{\href }%
\providecommand \doibase [0]{https://doi.org/}%
\providecommand \selectlanguage [0]{\@gobble}%
\providecommand \bibinfo  [0]{\@secondoftwo}%
\providecommand \bibfield  [0]{\@secondoftwo}%
\providecommand \translation [1]{[#1]}%
\providecommand \BibitemOpen [0]{}%
\providecommand \bibitemStop [0]{}%
\providecommand \bibitemNoStop [0]{.\EOS\space}%
\providecommand \EOS [0]{\spacefactor3000\relax}%
\providecommand \BibitemShut  [1]{\csname bibitem#1\endcsname}%
\let\auto@bib@innerbib\@empty
%</preamble>
\bibitem [{\citenamefont {RASHBA}(1960)}]{Rashba}%
  \BibitemOpen
  \bibfield  {author} {\bibinfo {author} {\bibfnamefont {E.}~\bibnamefont
  {RASHBA}},\ }\bibfield  {title} {\bibinfo {title} {Properties of
  semiconductors with an extremum loop. i. cyclotron and combinational
  resonance in a magnetic field perpendicular to the plane of the loop},\
  }\href {https://cir.nii.ac.jp/crid/1571698600346713472} {\bibfield  {journal}
  {\bibinfo  {journal} {Sov. Phys.-Solid State}\ }\textbf {\bibinfo {volume}
  {2}},\ \bibinfo {pages} {1109} (\bibinfo {year} {1960})}\BibitemShut
  {NoStop}%
\bibitem [{\citenamefont {Dresselhaus}(1955)}]{Dresselhaus1955}%
  \BibitemOpen
  \bibfield  {author} {\bibinfo {author} {\bibfnamefont {G.}~\bibnamefont
  {Dresselhaus}},\ }\bibfield  {title} {\bibinfo {title} {Spin-orbit coupling
  effects in zinc blende structures},\ }\href
  {https://doi.org/10.1103/physrev.100.580} {\bibfield  {journal} {\bibinfo
  {journal} {Phys. Rev.}\ }\textbf {\bibinfo {volume} {100}},\ \bibinfo {pages}
  {580} (\bibinfo {year} {1955})}\BibitemShut {NoStop}%
\bibitem [{\citenamefont {LaShell}\ \emph {et~al.}(1996)\citenamefont
  {LaShell}, \citenamefont {McDougall},\ and\ \citenamefont
  {Jensen}}]{LaShell1996}%
  \BibitemOpen
  \bibfield  {author} {\bibinfo {author} {\bibfnamefont {S.}~\bibnamefont
  {LaShell}}, \bibinfo {author} {\bibfnamefont {B.~A.}\ \bibnamefont
  {McDougall}},\ and\ \bibinfo {author} {\bibfnamefont {E.}~\bibnamefont
  {Jensen}},\ }\bibfield  {title} {\bibinfo {title} {Spin splitting of an
  au(111) surface state band observed with angle resolved photoelectron
  spectroscopy},\ }\href {https://doi.org/10.1103/physrevlett.77.3419}
  {\bibfield  {journal} {\bibinfo  {journal} {Phys. Rev. Lett.}\ }\textbf
  {\bibinfo {volume} {77}},\ \bibinfo {pages} {3419} (\bibinfo {year}
  {1996})}\BibitemShut {NoStop}%
\bibitem [{\citenamefont {Manchon}\ \emph {et~al.}(2015)\citenamefont
  {Manchon}, \citenamefont {Koo}, \citenamefont {Nitta}, \citenamefont
  {Frolov},\ and\ \citenamefont {Duine}}]{Manchon2015}%
  \BibitemOpen
  \bibfield  {author} {\bibinfo {author} {\bibfnamefont {A.}~\bibnamefont
  {Manchon}}, \bibinfo {author} {\bibfnamefont {H.~C.}\ \bibnamefont {Koo}},
  \bibinfo {author} {\bibfnamefont {J.}~\bibnamefont {Nitta}}, \bibinfo
  {author} {\bibfnamefont {S.~M.}\ \bibnamefont {Frolov}},\ and\ \bibinfo
  {author} {\bibfnamefont {R.~A.}\ \bibnamefont {Duine}},\ }\bibfield  {title}
  {\bibinfo {title} {New perspectives for rashba spin–orbit coupling},\
  }\href {https://doi.org/10.1038/nmat4360} {\bibfield  {journal} {\bibinfo
  {journal} {Nat. Mater.}\ }\textbf {\bibinfo {volume} {14}},\ \bibinfo {pages}
  {871} (\bibinfo {year} {2015})}\BibitemShut {NoStop}%
\bibitem [{\citenamefont {Kittel}(2004)}]{Kittel}%
  \BibitemOpen
  \bibfield  {author} {\bibinfo {author} {\bibfnamefont {C.}~\bibnamefont
  {Kittel}},\ }\href@noop {} {\emph {\bibinfo {title} {Introduction to Solid
  State Physics}}}\ (\bibinfo  {publisher} {8th Edition, John Wiley \& Sons,
  Inc., Hoboken.},\ \bibinfo {year} {2004})\BibitemShut {NoStop}%
\bibitem [{\citenamefont {Šmejkal}\ \emph
  {et~al.}(2022{\natexlab{a}})\citenamefont {Šmejkal}, \citenamefont
  {Sinova},\ and\ \citenamefont {Jungwirth}}]{Smejkal2022}%
  \BibitemOpen
  \bibfield  {author} {\bibinfo {author} {\bibfnamefont {L.}~\bibnamefont
  {Šmejkal}}, \bibinfo {author} {\bibfnamefont {J.}~\bibnamefont {Sinova}},\
  and\ \bibinfo {author} {\bibfnamefont {T.}~\bibnamefont {Jungwirth}},\
  }\bibfield  {title} {\bibinfo {title} {Beyond conventional ferromagnetism and
  antiferromagnetism: A phase with nonrelativistic spin and crystal rotation
  symmetry},\ }\href {https://doi.org/10.1103/physrevx.12.031042} {\bibfield
  {journal} {\bibinfo  {journal} {Phys. Rev. X}\ }\textbf {\bibinfo {volume}
  {12}},\ \bibinfo {pages} {031042} (\bibinfo {year}
  {2022}{\natexlab{a}})}\BibitemShut {NoStop}%
\bibitem [{\citenamefont {Šmejkal}\ \emph
  {et~al.}(2022{\natexlab{b}})\citenamefont {Šmejkal}, \citenamefont
  {Sinova},\ and\ \citenamefont {Jungwirth}}]{Smejkal2022a}%
  \BibitemOpen
  \bibfield  {author} {\bibinfo {author} {\bibfnamefont {L.}~\bibnamefont
  {Šmejkal}}, \bibinfo {author} {\bibfnamefont {J.}~\bibnamefont {Sinova}},\
  and\ \bibinfo {author} {\bibfnamefont {T.}~\bibnamefont {Jungwirth}},\
  }\bibfield  {title} {\bibinfo {title} {Emerging research landscape of
  altermagnetism},\ }\href {https://doi.org/10.1103/physrevx.12.040501}
  {\bibfield  {journal} {\bibinfo  {journal} {Phys. Rev. X}\ }\textbf {\bibinfo
  {volume} {12}},\ \bibinfo {pages} {040501} (\bibinfo {year}
  {2022}{\natexlab{b}})}\BibitemShut {NoStop}%
\bibitem [{\citenamefont {Mazin}(2022)}]{Mazin2022}%
  \BibitemOpen
  \bibfield  {author} {\bibinfo {author} {\bibfnamefont {I.~a.}\ \bibnamefont
  {Mazin}},\ }\bibfield  {title} {\bibinfo {title} {Editorial:
  Altermagnetism—a new punch line of fundamental magnetism},\ }\href
  {https://doi.org/10.1103/physrevx.12.040002} {\bibfield  {journal} {\bibinfo
  {journal} {Phys. Rev. X}\ }\textbf {\bibinfo {volume} {12}},\ \bibinfo
  {pages} {040002} (\bibinfo {year} {2022})}\BibitemShut {NoStop}%
\bibitem [{\citenamefont {Bai}\ \emph {et~al.}(2024)\citenamefont {Bai},
  \citenamefont {Feng}, \citenamefont {Liu}, \citenamefont {Šmejkal},
  \citenamefont {Mokrousov},\ and\ \citenamefont {Yao}}]{Bai2024}%
  \BibitemOpen
  \bibfield  {author} {\bibinfo {author} {\bibfnamefont {L.}~\bibnamefont
  {Bai}}, \bibinfo {author} {\bibfnamefont {W.}~\bibnamefont {Feng}}, \bibinfo
  {author} {\bibfnamefont {S.}~\bibnamefont {Liu}}, \bibinfo {author}
  {\bibfnamefont {L.}~\bibnamefont {Šmejkal}}, \bibinfo {author}
  {\bibfnamefont {Y.}~\bibnamefont {Mokrousov}},\ and\ \bibinfo {author}
  {\bibfnamefont {Y.}~\bibnamefont {Yao}},\ }\bibfield  {title} {\bibinfo
  {title} {Altermagnetism: Exploring new frontiers in magnetism and
  spintronics},\ }\href {https://doi.org/10.1002/adfm.202409327} {\bibfield
  {journal} {\bibinfo  {journal} {Adv. Funct. Mater.}\ }\textbf {\bibinfo
  {volume} {34}},\ \bibinfo {pages} {2409327} (\bibinfo {year}
  {2024})}\BibitemShut {NoStop}%
\bibitem [{Zha()}]{Zhang2024}%
  \BibitemOpen
  \href@noop {} {\ }\BibitemShut {NoStop}%
\bibitem [{\citenamefont {Guo}\ \emph {et~al.}(2025)\citenamefont {Guo},
  \citenamefont {Zong}, \citenamefont {Zhang}, \citenamefont {Wang},
  \citenamefont {Liu},\ and\ \citenamefont {Ji}}]{Guo2025}%
  \BibitemOpen
  \bibfield  {author} {\bibinfo {author} {\bibfnamefont {D.}~\bibnamefont
  {Guo}}, \bibinfo {author} {\bibfnamefont {C.}~\bibnamefont {Zong}}, \bibinfo
  {author} {\bibfnamefont {W.}~\bibnamefont {Zhang}}, \bibinfo {author}
  {\bibfnamefont {C.}~\bibnamefont {Wang}}, \bibinfo {author} {\bibfnamefont
  {J.}~\bibnamefont {Liu}},\ and\ \bibinfo {author} {\bibfnamefont
  {W.}~\bibnamefont {Ji}},\ }\bibfield  {title} {\bibinfo {title} {Tunable
  altermagnetism via interchain engineering in parallel-assembled atomic
  chains},\ }\href {https://doi.org/10.1103/zhds-vnyt} {\bibfield  {journal}
  {\bibinfo  {journal} {Phys. Rev. B}\ }\textbf {\bibinfo {volume} {112}},\
  \bibinfo {pages} {L041404} (\bibinfo {year} {2025})}\BibitemShut {NoStop}%
\bibitem [{\citenamefont {Huang}\ \emph {et~al.}(2026)\citenamefont {Huang},
  \citenamefont {Qin}, \citenamefont {Zhan}, \citenamefont {Xu}, \citenamefont
  {Ma},\ and\ \citenamefont {Wang}}]{Huang2026}%
  \BibitemOpen
  \bibfield  {author} {\bibinfo {author} {\bibfnamefont {S.}~\bibnamefont
  {Huang}}, \bibinfo {author} {\bibfnamefont {Z.}~\bibnamefont {Qin}}, \bibinfo
  {author} {\bibfnamefont {F.}~\bibnamefont {Zhan}}, \bibinfo {author}
  {\bibfnamefont {D.-H.}\ \bibnamefont {Xu}}, \bibinfo {author} {\bibfnamefont
  {D.-S.}\ \bibnamefont {Ma}},\ and\ \bibinfo {author} {\bibfnamefont
  {R.}~\bibnamefont {Wang}},\ }\bibfield  {title} {\bibinfo {title}
  {Light-induced odd-parity magnetism in conventional antiferromagnetism},\
  }\href {https://doi.org/10.1103/9346-9jpf} {\bibfield  {journal} {\bibinfo
  {journal} {Phys. Rev. Lett.}\ }\textbf {\bibinfo {volume} {136}},\ \bibinfo
  {pages} {126703} (\bibinfo {year} {2026})}\BibitemShut {NoStop}%
\bibitem [{\citenamefont {Zhu}\ \emph {et~al.}(2026)\citenamefont {Zhu},
  \citenamefont {Zhou}, \citenamefont {Wang}, \citenamefont {Wei},\ and\
  \citenamefont {Ruan}}]{Zhu2026}%
  \BibitemOpen
  \bibfield  {author} {\bibinfo {author} {\bibfnamefont {T.}~\bibnamefont
  {Zhu}}, \bibinfo {author} {\bibfnamefont {D.}~\bibnamefont {Zhou}}, \bibinfo
  {author} {\bibfnamefont {H.}~\bibnamefont {Wang}}, \bibinfo {author}
  {\bibfnamefont {S.-H.}\ \bibnamefont {Wei}},\ and\ \bibinfo {author}
  {\bibfnamefont {J.}~\bibnamefont {Ruan}},\ }\bibfield  {title} {\bibinfo
  {title} {Floquet odd-parity collinear magnets},\ }\href
  {https://doi.org/10.1103/7ywb-ml2q} {\bibfield  {journal} {\bibinfo
  {journal} {Phys. Rev. Lett.}\ }\textbf {\bibinfo {volume} {136}},\ \bibinfo
  {pages} {126704} (\bibinfo {year} {2026})}\BibitemShut {NoStop}%
\bibitem [{\citenamefont {Hayami}\ \emph {et~al.}(2019)\citenamefont {Hayami},
  \citenamefont {Yanagi},\ and\ \citenamefont {Kusunose}}]{Hayami2019}%
  \BibitemOpen
  \bibfield  {author} {\bibinfo {author} {\bibfnamefont {S.}~\bibnamefont
  {Hayami}}, \bibinfo {author} {\bibfnamefont {Y.}~\bibnamefont {Yanagi}},\
  and\ \bibinfo {author} {\bibfnamefont {H.}~\bibnamefont {Kusunose}},\
  }\bibfield  {title} {\bibinfo {title} {Momentum-dependent spin splitting by
  collinear antiferromagnetic ordering},\ }\href
  {https://doi.org/10.7566/jpsj.88.123702} {\bibfield  {journal} {\bibinfo
  {journal} {J. Phys. Soc. Japan}\ }\textbf {\bibinfo {volume} {88}},\ \bibinfo
  {pages} {123702} (\bibinfo {year} {2019})}\BibitemShut {NoStop}%
\bibitem [{\citenamefont {Ahn}\ \emph {et~al.}(2019)\citenamefont {Ahn},
  \citenamefont {Hariki}, \citenamefont {Lee},\ and\ \citenamefont
  {Kuneš}}]{Ahn2019}%
  \BibitemOpen
  \bibfield  {author} {\bibinfo {author} {\bibfnamefont {K.-H.}\ \bibnamefont
  {Ahn}}, \bibinfo {author} {\bibfnamefont {A.}~\bibnamefont {Hariki}},
  \bibinfo {author} {\bibfnamefont {K.-W.}\ \bibnamefont {Lee}},\ and\ \bibinfo
  {author} {\bibfnamefont {J.}~\bibnamefont {Kuneš}},\ }\bibfield  {title}
  {\bibinfo {title} {Antiferromagnetism in ruo2 as d -wave pomeranchuk
  instability},\ }\href {https://doi.org/10.1103/physrevb.99.184432} {\bibfield
   {journal} {\bibinfo  {journal} {Phys. Rev. B}\ }\textbf {\bibinfo {volume}
  {99}},\ \bibinfo {pages} {184432} (\bibinfo {year} {2019})}\BibitemShut
  {NoStop}%
\bibitem [{\citenamefont {González-Hernández}\ \emph
  {et~al.}(2021)\citenamefont {González-Hernández}, \citenamefont {Šmejkal},
  \citenamefont {Výborný}, \citenamefont {Yahagi}, \citenamefont {Sinova},
  \citenamefont {Jungwirth},\ and\ \citenamefont
  {Železný}}]{GonzalezHernandez2021}%
  \BibitemOpen
  \bibfield  {author} {\bibinfo {author} {\bibfnamefont {R.}~\bibnamefont
  {González-Hernández}}, \bibinfo {author} {\bibfnamefont {L.}~\bibnamefont
  {Šmejkal}}, \bibinfo {author} {\bibfnamefont {K.}~\bibnamefont {Výborný}},
  \bibinfo {author} {\bibfnamefont {Y.}~\bibnamefont {Yahagi}}, \bibinfo
  {author} {\bibfnamefont {J.}~\bibnamefont {Sinova}}, \bibinfo {author}
  {\bibfnamefont {T.}~\bibnamefont {Jungwirth}},\ and\ \bibinfo {author}
  {\bibfnamefont {J.}~\bibnamefont {Železný}},\ }\bibfield  {title} {\bibinfo
  {title} {Efficient electrical spin splitter based on nonrelativistic
  collinear antiferromagnetism},\ }\href
  {https://doi.org/10.1103/physrevlett.126.127701} {\bibfield  {journal}
  {\bibinfo  {journal} {Phys. Rev. Lett.}\ }\textbf {\bibinfo {volume} {126}},\
  \bibinfo {pages} {127701} (\bibinfo {year} {2021})}\BibitemShut {NoStop}%
\bibitem [{\citenamefont {Leeb}\ \emph {et~al.}(2024)\citenamefont {Leeb},
  \citenamefont {Mook}, \citenamefont {Šmejkal},\ and\ \citenamefont
  {Knolle}}]{Leeb2024}%
  \BibitemOpen
  \bibfield  {author} {\bibinfo {author} {\bibfnamefont {V.}~\bibnamefont
  {Leeb}}, \bibinfo {author} {\bibfnamefont {A.}~\bibnamefont {Mook}}, \bibinfo
  {author} {\bibfnamefont {L.}~\bibnamefont {Šmejkal}},\ and\ \bibinfo
  {author} {\bibfnamefont {J.}~\bibnamefont {Knolle}},\ }\bibfield  {title}
  {\bibinfo {title} {Spontaneous formation of altermagnetism from orbital
  ordering},\ }\href {https://doi.org/10.1103/physrevlett.132.236701}
  {\bibfield  {journal} {\bibinfo  {journal} {Phys. Rev. Lett.}\ }\textbf
  {\bibinfo {volume} {132}},\ \bibinfo {pages} {236701} (\bibinfo {year}
  {2024})}\BibitemShut {NoStop}%
\bibitem [{\citenamefont {Krempaský}\ \emph {et~al.}(2024)\citenamefont
  {Krempaský}, \citenamefont {Šmejkal}, \citenamefont {D’Souza},
  \citenamefont {Hajlaoui}, \citenamefont {Springholz}, \citenamefont
  {Uhlířová}, \citenamefont {Alarab}, \citenamefont {Constantinou},
  \citenamefont {Strocov}, \citenamefont {Usanov}, \citenamefont {Pudelko},
  \citenamefont {González-Hernández}, \citenamefont {Birk~Hellenes},
  \citenamefont {Jansa}, \citenamefont {Reichlová}, \citenamefont {Šobáň},
  \citenamefont {Gonzalez~Betancourt}, \citenamefont {Wadley}, \citenamefont
  {Sinova}, \citenamefont {Kriegner}, \citenamefont {Minár}, \citenamefont
  {Dil},\ and\ \citenamefont {Jungwirth}}]{Krempasky2024}%
  \BibitemOpen
  \bibfield  {author} {\bibinfo {author} {\bibfnamefont {J.}~\bibnamefont
  {Krempaský}}, \bibinfo {author} {\bibfnamefont {L.}~\bibnamefont
  {Šmejkal}}, \bibinfo {author} {\bibfnamefont {S.~W.}\ \bibnamefont
  {D’Souza}}, \bibinfo {author} {\bibfnamefont {M.}~\bibnamefont {Hajlaoui}},
  \bibinfo {author} {\bibfnamefont {G.}~\bibnamefont {Springholz}}, \bibinfo
  {author} {\bibfnamefont {K.}~\bibnamefont {Uhlířová}}, \bibinfo {author}
  {\bibfnamefont {F.}~\bibnamefont {Alarab}}, \bibinfo {author} {\bibfnamefont
  {P.~C.}\ \bibnamefont {Constantinou}}, \bibinfo {author} {\bibfnamefont
  {V.}~\bibnamefont {Strocov}}, \bibinfo {author} {\bibfnamefont
  {D.}~\bibnamefont {Usanov}}, \bibinfo {author} {\bibfnamefont {W.~R.}\
  \bibnamefont {Pudelko}}, \bibinfo {author} {\bibfnamefont {R.}~\bibnamefont
  {González-Hernández}}, \bibinfo {author} {\bibfnamefont {A.}~\bibnamefont
  {Birk~Hellenes}}, \bibinfo {author} {\bibfnamefont {Z.}~\bibnamefont
  {Jansa}}, \bibinfo {author} {\bibfnamefont {H.}~\bibnamefont {Reichlová}},
  \bibinfo {author} {\bibfnamefont {Z.}~\bibnamefont {Šobáň}}, \bibinfo
  {author} {\bibfnamefont {R.~D.}\ \bibnamefont {Gonzalez~Betancourt}},
  \bibinfo {author} {\bibfnamefont {P.}~\bibnamefont {Wadley}}, \bibinfo
  {author} {\bibfnamefont {J.}~\bibnamefont {Sinova}}, \bibinfo {author}
  {\bibfnamefont {D.}~\bibnamefont {Kriegner}}, \bibinfo {author}
  {\bibfnamefont {J.}~\bibnamefont {Minár}}, \bibinfo {author} {\bibfnamefont
  {J.~H.}\ \bibnamefont {Dil}},\ and\ \bibinfo {author} {\bibfnamefont
  {T.}~\bibnamefont {Jungwirth}},\ }\bibfield  {title} {\bibinfo {title}
  {Altermagnetic lifting of kramers spin degeneracy},\ }\href
  {https://doi.org/10.1038/s41586-023-06907-7} {\bibfield  {journal} {\bibinfo
  {journal} {Nature}\ }\textbf {\bibinfo {volume} {626}},\ \bibinfo {pages}
  {517} (\bibinfo {year} {2024})}\BibitemShut {NoStop}%
\bibitem [{\citenamefont {Lee}\ \emph {et~al.}(2024)\citenamefont {Lee},
  \citenamefont {Lee}, \citenamefont {Jung}, \citenamefont {Jung},
  \citenamefont {Kim}, \citenamefont {Lee}, \citenamefont {Seok}, \citenamefont
  {Kim}, \citenamefont {Park}, \citenamefont {Šmejkal}, \citenamefont {Kang},\
  and\ \citenamefont {Kim}}]{Lee2024}%
  \BibitemOpen
  \bibfield  {author} {\bibinfo {author} {\bibfnamefont {S.}~\bibnamefont
  {Lee}}, \bibinfo {author} {\bibfnamefont {S.}~\bibnamefont {Lee}}, \bibinfo
  {author} {\bibfnamefont {S.}~\bibnamefont {Jung}}, \bibinfo {author}
  {\bibfnamefont {J.}~\bibnamefont {Jung}}, \bibinfo {author} {\bibfnamefont
  {D.}~\bibnamefont {Kim}}, \bibinfo {author} {\bibfnamefont {Y.}~\bibnamefont
  {Lee}}, \bibinfo {author} {\bibfnamefont {B.}~\bibnamefont {Seok}}, \bibinfo
  {author} {\bibfnamefont {J.}~\bibnamefont {Kim}}, \bibinfo {author}
  {\bibfnamefont {B.~G.}\ \bibnamefont {Park}}, \bibinfo {author}
  {\bibfnamefont {L.}~\bibnamefont {Šmejkal}}, \bibinfo {author}
  {\bibfnamefont {C.-J.}\ \bibnamefont {Kang}},\ and\ \bibinfo {author}
  {\bibfnamefont {C.}~\bibnamefont {Kim}},\ }\bibfield  {title} {\bibinfo
  {title} {Broken kramers degeneracy in altermagnetic mnte},\ }\href
  {https://doi.org/10.1103/physrevlett.132.036702} {\bibfield  {journal}
  {\bibinfo  {journal} {Phys. Rev. Lett.}\ }\textbf {\bibinfo {volume} {132}},\
  \bibinfo {pages} {036702} (\bibinfo {year} {2024})}\BibitemShut {NoStop}%
\bibitem [{\citenamefont {Reimers}\ \emph {et~al.}(2024)\citenamefont
  {Reimers}, \citenamefont {Odenbreit}, \citenamefont {Šmejkal}, \citenamefont
  {Strocov}, \citenamefont {Constantinou}, \citenamefont {Hellenes},
  \citenamefont {Jaeschke~Ubiergo}, \citenamefont {Campos}, \citenamefont
  {Bharadwaj}, \citenamefont {Chakraborty}, \citenamefont {Denneulin},
  \citenamefont {Shi}, \citenamefont {Dunin-Borkowski}, \citenamefont {Das},
  \citenamefont {Kläui}, \citenamefont {Sinova},\ and\ \citenamefont
  {Jourdan}}]{Reimers2024}%
  \BibitemOpen
  \bibfield  {author} {\bibinfo {author} {\bibfnamefont {S.}~\bibnamefont
  {Reimers}}, \bibinfo {author} {\bibfnamefont {L.}~\bibnamefont {Odenbreit}},
  \bibinfo {author} {\bibfnamefont {L.}~\bibnamefont {Šmejkal}}, \bibinfo
  {author} {\bibfnamefont {V.~N.}\ \bibnamefont {Strocov}}, \bibinfo {author}
  {\bibfnamefont {P.}~\bibnamefont {Constantinou}}, \bibinfo {author}
  {\bibfnamefont {A.~B.}\ \bibnamefont {Hellenes}}, \bibinfo {author}
  {\bibfnamefont {R.}~\bibnamefont {Jaeschke~Ubiergo}}, \bibinfo {author}
  {\bibfnamefont {W.~H.}\ \bibnamefont {Campos}}, \bibinfo {author}
  {\bibfnamefont {V.~K.}\ \bibnamefont {Bharadwaj}}, \bibinfo {author}
  {\bibfnamefont {A.}~\bibnamefont {Chakraborty}}, \bibinfo {author}
  {\bibfnamefont {T.}~\bibnamefont {Denneulin}}, \bibinfo {author}
  {\bibfnamefont {W.}~\bibnamefont {Shi}}, \bibinfo {author} {\bibfnamefont
  {R.~E.}\ \bibnamefont {Dunin-Borkowski}}, \bibinfo {author} {\bibfnamefont
  {S.}~\bibnamefont {Das}}, \bibinfo {author} {\bibfnamefont {M.}~\bibnamefont
  {Kläui}}, \bibinfo {author} {\bibfnamefont {J.}~\bibnamefont {Sinova}},\
  and\ \bibinfo {author} {\bibfnamefont {M.}~\bibnamefont {Jourdan}},\
  }\bibfield  {title} {\bibinfo {title} {Direct observation of altermagnetic
  band splitting in crsb thin films},\ }\href
  {https://doi.org/10.1038/s41467-024-46476-5} {\bibfield  {journal} {\bibinfo
  {journal} {Nat. Commun.}\ }\textbf {\bibinfo {volume} {15}},\ \bibinfo
  {pages} {2116} (\bibinfo {year} {2024})}\BibitemShut {NoStop}%
\bibitem [{\citenamefont {Jiang}\ \emph {et~al.}(2025)\citenamefont {Jiang},
  \citenamefont {Hu}, \citenamefont {Bai}, \citenamefont {Song}, \citenamefont
  {Mu}, \citenamefont {Qu}, \citenamefont {Li}, \citenamefont {Zhu},
  \citenamefont {Pi}, \citenamefont {Wei}, \citenamefont {Sun}, \citenamefont
  {Huang}, \citenamefont {Zheng}, \citenamefont {Peng}, \citenamefont {He},
  \citenamefont {Li}, \citenamefont {Luo}, \citenamefont {Li}, \citenamefont
  {Chen}, \citenamefont {Li}, \citenamefont {Weng},\ and\ \citenamefont
  {Qian}}]{Jiang2025}%
  \BibitemOpen
  \bibfield  {author} {\bibinfo {author} {\bibfnamefont {B.}~\bibnamefont
  {Jiang}}, \bibinfo {author} {\bibfnamefont {M.}~\bibnamefont {Hu}}, \bibinfo
  {author} {\bibfnamefont {J.}~\bibnamefont {Bai}}, \bibinfo {author}
  {\bibfnamefont {Z.}~\bibnamefont {Song}}, \bibinfo {author} {\bibfnamefont
  {C.}~\bibnamefont {Mu}}, \bibinfo {author} {\bibfnamefont {G.}~\bibnamefont
  {Qu}}, \bibinfo {author} {\bibfnamefont {W.}~\bibnamefont {Li}}, \bibinfo
  {author} {\bibfnamefont {W.}~\bibnamefont {Zhu}}, \bibinfo {author}
  {\bibfnamefont {H.}~\bibnamefont {Pi}}, \bibinfo {author} {\bibfnamefont
  {Z.}~\bibnamefont {Wei}}, \bibinfo {author} {\bibfnamefont {Y.-J.}\
  \bibnamefont {Sun}}, \bibinfo {author} {\bibfnamefont {Y.}~\bibnamefont
  {Huang}}, \bibinfo {author} {\bibfnamefont {X.}~\bibnamefont {Zheng}},
  \bibinfo {author} {\bibfnamefont {Y.}~\bibnamefont {Peng}}, \bibinfo {author}
  {\bibfnamefont {L.}~\bibnamefont {He}}, \bibinfo {author} {\bibfnamefont
  {S.}~\bibnamefont {Li}}, \bibinfo {author} {\bibfnamefont {J.}~\bibnamefont
  {Luo}}, \bibinfo {author} {\bibfnamefont {Z.}~\bibnamefont {Li}}, \bibinfo
  {author} {\bibfnamefont {G.}~\bibnamefont {Chen}}, \bibinfo {author}
  {\bibfnamefont {H.}~\bibnamefont {Li}}, \bibinfo {author} {\bibfnamefont
  {H.}~\bibnamefont {Weng}},\ and\ \bibinfo {author} {\bibfnamefont
  {T.}~\bibnamefont {Qian}},\ }\bibfield  {title} {\bibinfo {title} {A metallic
  room-temperature d-wave altermagnet},\ }\href
  {https://doi.org/10.1038/s41567-025-02822-y} {\bibfield  {journal} {\bibinfo
  {journal} {Nat. Phys.}\ }\textbf {\bibinfo {volume} {21}},\ \bibinfo {pages}
  {754} (\bibinfo {year} {2025})}\BibitemShut {NoStop}%
\bibitem [{\citenamefont {Cao}\ \emph {et~al.}(2025)\citenamefont {Cao},
  \citenamefont {Denisov}, \citenamefont {Liu},\ and\ \citenamefont
  {Žutić}}]{Cao2025}%
  \BibitemOpen
  \bibfield  {author} {\bibinfo {author} {\bibfnamefont {J.~D.}\ \bibnamefont
  {Cao}}, \bibinfo {author} {\bibfnamefont {K.~S.}\ \bibnamefont {Denisov}},
  \bibinfo {author} {\bibfnamefont {Y.}~\bibnamefont {Liu}},\ and\ \bibinfo
  {author} {\bibfnamefont {I.}~\bibnamefont {Žutić}},\ }\bibfield  {title}
  {\bibinfo {title} {Symmetry classification for alternating excitons in
  two-dimensional altermagnets},\ }\href {https://doi.org/10.1103/zn7r-k1xd}
  {\bibfield  {journal} {\bibinfo  {journal} {Phys. Rev. Lett.}\ }\textbf
  {\bibinfo {volume} {135}},\ \bibinfo {pages} {266703} (\bibinfo {year}
  {2025})}\BibitemShut {NoStop}%
\bibitem [{\citenamefont {Zhang}\ \emph {et~al.}(2018)\citenamefont {Zhang},
  \citenamefont {Song}, \citenamefont {Alexandradinata}, \citenamefont {Weng},
  \citenamefont {Fang}, \citenamefont {Lu},\ and\ \citenamefont
  {Fang}}]{Zhang2018}%
  \BibitemOpen
  \bibfield  {author} {\bibinfo {author} {\bibfnamefont {T.}~\bibnamefont
  {Zhang}}, \bibinfo {author} {\bibfnamefont {Z.}~\bibnamefont {Song}},
  \bibinfo {author} {\bibfnamefont {A.}~\bibnamefont {Alexandradinata}},
  \bibinfo {author} {\bibfnamefont {H.}~\bibnamefont {Weng}}, \bibinfo {author}
  {\bibfnamefont {C.}~\bibnamefont {Fang}}, \bibinfo {author} {\bibfnamefont
  {L.}~\bibnamefont {Lu}},\ and\ \bibinfo {author} {\bibfnamefont
  {Z.}~\bibnamefont {Fang}},\ }\bibfield  {title} {\bibinfo {title}
  {Double-weyl phonons in transition-metal monosilicides},\ }\href
  {https://doi.org/10.1103/physrevlett.120.016401} {\bibfield  {journal}
  {\bibinfo  {journal} {Phys. Rev. Lett.}\ }\textbf {\bibinfo {volume} {120}},\
  \bibinfo {pages} {016401} (\bibinfo {year} {2018})}\BibitemShut {NoStop}%
\bibitem [{\citenamefont {Jin}\ \emph {et~al.}(2018)\citenamefont {Jin},
  \citenamefont {Chen}, \citenamefont {Xia}, \citenamefont {Zhao},
  \citenamefont {Wang},\ and\ \citenamefont {Xu}}]{Jin2018}%
  \BibitemOpen
  \bibfield  {author} {\bibinfo {author} {\bibfnamefont {Y.~J.}\ \bibnamefont
  {Jin}}, \bibinfo {author} {\bibfnamefont {Z.~J.}\ \bibnamefont {Chen}},
  \bibinfo {author} {\bibfnamefont {B.~W.}\ \bibnamefont {Xia}}, \bibinfo
  {author} {\bibfnamefont {Y.~J.}\ \bibnamefont {Zhao}}, \bibinfo {author}
  {\bibfnamefont {R.}~\bibnamefont {Wang}},\ and\ \bibinfo {author}
  {\bibfnamefont {H.}~\bibnamefont {Xu}},\ }\bibfield  {title} {\bibinfo
  {title} {Ideal intersecting nodal-ring phonons in bcc c8},\ }\href
  {https://doi.org/10.1103/physrevb.98.220103} {\bibfield  {journal} {\bibinfo
  {journal} {Phys. Rev. B}\ }\textbf {\bibinfo {volume} {98}},\ \bibinfo
  {pages} {220103} (\bibinfo {year} {2018})}\BibitemShut {NoStop}%
\bibitem [{\citenamefont {Liu}\ \emph {et~al.}(2020)\citenamefont {Liu},
  \citenamefont {Chen},\ and\ \citenamefont {Xu}}]{Liu2019}%
  \BibitemOpen
  \bibfield  {author} {\bibinfo {author} {\bibfnamefont {Y.}~\bibnamefont
  {Liu}}, \bibinfo {author} {\bibfnamefont {X.}~\bibnamefont {Chen}},\ and\
  \bibinfo {author} {\bibfnamefont {Y.}~\bibnamefont {Xu}},\ }\bibfield
  {title} {\bibinfo {title} {Topological phononics: From fundamental models to
  real materials},\ }\href {https://doi.org/10.1002/adfm.201904784} {\bibfield
  {journal} {\bibinfo  {journal} {Adv. Funct. Mater.}\ }\textbf {\bibinfo
  {volume} {30}},\ \bibinfo {pages} {1904784} (\bibinfo {year}
  {2020})}\BibitemShut {NoStop}%
\bibitem [{\citenamefont {Xu}\ \emph {et~al.}(2024)\citenamefont {Xu},
  \citenamefont {Vergniory}, \citenamefont {Ma}, \citenamefont {Mañes},
  \citenamefont {Song}, \citenamefont {Bernevig}, \citenamefont {Regnault},\
  and\ \citenamefont {Elcoro}}]{Xu2024}%
  \BibitemOpen
  \bibfield  {author} {\bibinfo {author} {\bibfnamefont {Y.}~\bibnamefont
  {Xu}}, \bibinfo {author} {\bibfnamefont {M.~G.}\ \bibnamefont {Vergniory}},
  \bibinfo {author} {\bibfnamefont {D.-S.}\ \bibnamefont {Ma}}, \bibinfo
  {author} {\bibfnamefont {J.~L.}\ \bibnamefont {Mañes}}, \bibinfo {author}
  {\bibfnamefont {Z.-D.}\ \bibnamefont {Song}}, \bibinfo {author}
  {\bibfnamefont {B.~A.}\ \bibnamefont {Bernevig}}, \bibinfo {author}
  {\bibfnamefont {N.}~\bibnamefont {Regnault}},\ and\ \bibinfo {author}
  {\bibfnamefont {L.}~\bibnamefont {Elcoro}},\ }\bibfield  {title} {\bibinfo
  {title} {Catalog of topological phonon materials},\ }\href
  {https://doi.org/10.1126/science.adf8458} {\bibfield  {journal} {\bibinfo
  {journal} {Science}\ }\textbf {\bibinfo {volume} {384}},\ \bibinfo {pages}
  {638} (\bibinfo {year} {2024})}\BibitemShut {NoStop}%
\bibitem [{\citenamefont {Zhang}\ and\ \citenamefont {Niu}(2014)}]{Zhang2014}%
  \BibitemOpen
  \bibfield  {author} {\bibinfo {author} {\bibfnamefont {L.}~\bibnamefont
  {Zhang}}\ and\ \bibinfo {author} {\bibfnamefont {Q.}~\bibnamefont {Niu}},\
  }\bibfield  {title} {\bibinfo {title} {Angular momentum of phonons and the
  einstein–de haas effect},\ }\href
  {https://doi.org/10.1103/physrevlett.112.085503} {\bibfield  {journal}
  {\bibinfo  {journal} {Phys. Rev. Lett.}\ }\textbf {\bibinfo {volume} {112}},\
  \bibinfo {pages} {085503} (\bibinfo {year} {2014})}\BibitemShut {NoStop}%
\bibitem [{\citenamefont {Zhu}\ \emph {et~al.}(2018)\citenamefont {Zhu},
  \citenamefont {Yi}, \citenamefont {Li}, \citenamefont {Xiao}, \citenamefont
  {Zhang}, \citenamefont {Yang}, \citenamefont {Kaindl}, \citenamefont {Li},
  \citenamefont {Wang},\ and\ \citenamefont {Zhang}}]{Zhu2018}%
  \BibitemOpen
  \bibfield  {author} {\bibinfo {author} {\bibfnamefont {H.}~\bibnamefont
  {Zhu}}, \bibinfo {author} {\bibfnamefont {J.}~\bibnamefont {Yi}}, \bibinfo
  {author} {\bibfnamefont {M.-Y.}\ \bibnamefont {Li}}, \bibinfo {author}
  {\bibfnamefont {J.}~\bibnamefont {Xiao}}, \bibinfo {author} {\bibfnamefont
  {L.}~\bibnamefont {Zhang}}, \bibinfo {author} {\bibfnamefont {C.-W.}\
  \bibnamefont {Yang}}, \bibinfo {author} {\bibfnamefont {R.~A.}\ \bibnamefont
  {Kaindl}}, \bibinfo {author} {\bibfnamefont {L.-J.}\ \bibnamefont {Li}},
  \bibinfo {author} {\bibfnamefont {Y.}~\bibnamefont {Wang}},\ and\ \bibinfo
  {author} {\bibfnamefont {X.}~\bibnamefont {Zhang}},\ }\bibfield  {title}
  {\bibinfo {title} {Observation of chiral phonons},\ }\href
  {https://doi.org/10.1126/science.aar2711} {\bibfield  {journal} {\bibinfo
  {journal} {Science}\ }\textbf {\bibinfo {volume} {359}},\ \bibinfo {pages}
  {579} (\bibinfo {year} {2018})}\BibitemShut {NoStop}%
\bibitem [{\citenamefont {Ishito}\ \emph {et~al.}(2023)\citenamefont {Ishito},
  \citenamefont {Mao}, \citenamefont {Kousaka}, \citenamefont {Togawa},
  \citenamefont {Iwasaki}, \citenamefont {Zhang}, \citenamefont {Murakami},
  \citenamefont {Kishine},\ and\ \citenamefont {Satoh}}]{Ishito2022}%
  \BibitemOpen
  \bibfield  {author} {\bibinfo {author} {\bibfnamefont {K.}~\bibnamefont
  {Ishito}}, \bibinfo {author} {\bibfnamefont {H.}~\bibnamefont {Mao}},
  \bibinfo {author} {\bibfnamefont {Y.}~\bibnamefont {Kousaka}}, \bibinfo
  {author} {\bibfnamefont {Y.}~\bibnamefont {Togawa}}, \bibinfo {author}
  {\bibfnamefont {S.}~\bibnamefont {Iwasaki}}, \bibinfo {author} {\bibfnamefont
  {T.}~\bibnamefont {Zhang}}, \bibinfo {author} {\bibfnamefont
  {S.}~\bibnamefont {Murakami}}, \bibinfo {author} {\bibfnamefont {J.-i.}\
  \bibnamefont {Kishine}},\ and\ \bibinfo {author} {\bibfnamefont
  {T.}~\bibnamefont {Satoh}},\ }\bibfield  {title} {\bibinfo {title} {Truly
  chiral phonons in $\alpha$-hgs},\ }\href
  {https://doi.org/10.1038/s41567-022-01790-x} {\bibfield  {journal} {\bibinfo
  {journal} {Nat. Phys.}\ }\textbf {\bibinfo {volume} {19}},\ \bibinfo {pages}
  {35} (\bibinfo {year} {2023})}\BibitemShut {NoStop}%
\bibitem [{\citenamefont {Strohm}\ \emph {et~al.}(2005)\citenamefont {Strohm},
  \citenamefont {Rikken},\ and\ \citenamefont {Wyder}}]{Strohm2005}%
  \BibitemOpen
  \bibfield  {author} {\bibinfo {author} {\bibfnamefont {C.}~\bibnamefont
  {Strohm}}, \bibinfo {author} {\bibfnamefont {G.~L. J.~A.}\ \bibnamefont
  {Rikken}},\ and\ \bibinfo {author} {\bibfnamefont {P.}~\bibnamefont
  {Wyder}},\ }\bibfield  {title} {\bibinfo {title} {Phenomenological evidence
  for the phonon hall effect},\ }\href
  {https://doi.org/10.1103/physrevlett.95.155901} {\bibfield  {journal}
  {\bibinfo  {journal} {Phys. Rev. Lett.}\ }\textbf {\bibinfo {volume} {95}},\
  \bibinfo {pages} {155901} (\bibinfo {year} {2005})}\BibitemShut {NoStop}%
\bibitem [{\citenamefont {Sheng}\ \emph {et~al.}(2006)\citenamefont {Sheng},
  \citenamefont {Sheng},\ and\ \citenamefont {Ting}}]{Sheng2006}%
  \BibitemOpen
  \bibfield  {author} {\bibinfo {author} {\bibfnamefont {L.}~\bibnamefont
  {Sheng}}, \bibinfo {author} {\bibfnamefont {D.~N.}\ \bibnamefont {Sheng}},\
  and\ \bibinfo {author} {\bibfnamefont {C.~S.}\ \bibnamefont {Ting}},\
  }\bibfield  {title} {\bibinfo {title} {Theory of the phonon hall effect in
  paramagnetic dielectrics},\ }\href
  {https://doi.org/10.1103/physrevlett.96.155901} {\bibfield  {journal}
  {\bibinfo  {journal} {Phys. Rev. Lett.}\ }\textbf {\bibinfo {volume} {96}},\
  \bibinfo {pages} {155901} (\bibinfo {year} {2006})}\BibitemShut {NoStop}%
\bibitem [{\citenamefont {Mori}\ \emph {et~al.}(2014)\citenamefont {Mori},
  \citenamefont {Spencer-Smith}, \citenamefont {Sushkov},\ and\ \citenamefont
  {Maekawa}}]{Mori2014}%
  \BibitemOpen
  \bibfield  {author} {\bibinfo {author} {\bibfnamefont {M.}~\bibnamefont
  {Mori}}, \bibinfo {author} {\bibfnamefont {A.}~\bibnamefont {Spencer-Smith}},
  \bibinfo {author} {\bibfnamefont {O.~P.}\ \bibnamefont {Sushkov}},\ and\
  \bibinfo {author} {\bibfnamefont {S.}~\bibnamefont {Maekawa}},\ }\bibfield
  {title} {\bibinfo {title} {Origin of the phonon hall effect in rare-earth
  garnets},\ }\href {https://doi.org/10.1103/physrevlett.113.265901} {\bibfield
   {journal} {\bibinfo  {journal} {Phys. Rev. Lett.}\ }\textbf {\bibinfo
  {volume} {113}},\ \bibinfo {pages} {265901} (\bibinfo {year}
  {2014})}\BibitemShut {NoStop}%
\bibitem [{\citenamefont {Li}\ \emph {et~al.}(2020)\citenamefont {Li},
  \citenamefont {Fauqué}, \citenamefont {Zhu},\ and\ \citenamefont
  {Behnia}}]{Li2020}%
  \BibitemOpen
  \bibfield  {author} {\bibinfo {author} {\bibfnamefont {X.}~\bibnamefont
  {Li}}, \bibinfo {author} {\bibfnamefont {B.}~\bibnamefont {Fauqué}},
  \bibinfo {author} {\bibfnamefont {Z.}~\bibnamefont {Zhu}},\ and\ \bibinfo
  {author} {\bibfnamefont {K.}~\bibnamefont {Behnia}},\ }\bibfield  {title}
  {\bibinfo {title} {Phonon thermal hall effect in strontium titanate},\ }\href
  {https://doi.org/10.1103/physrevlett.124.105901} {\bibfield  {journal}
  {\bibinfo  {journal} {Phys. Rev. Lett.}\ }\textbf {\bibinfo {volume} {124}},\
  \bibinfo {pages} {105901} (\bibinfo {year} {2020})}\BibitemShut {NoStop}%
\bibitem [{\citenamefont {Grissonnanche}\ \emph {et~al.}(2020)\citenamefont
  {Grissonnanche}, \citenamefont {Thériault}, \citenamefont {Gourgout},
  \citenamefont {Boulanger}, \citenamefont {Lefrançois}, \citenamefont
  {Ataei}, \citenamefont {Laliberté}, \citenamefont {Dion}, \citenamefont
  {Zhou}, \citenamefont {Pyon}, \citenamefont {Takayama}, \citenamefont
  {Takagi}, \citenamefont {Doiron-Leyraud},\ and\ \citenamefont
  {Taillefer}}]{Grissonnanche2020}%
  \BibitemOpen
  \bibfield  {author} {\bibinfo {author} {\bibfnamefont {G.}~\bibnamefont
  {Grissonnanche}}, \bibinfo {author} {\bibfnamefont {S.}~\bibnamefont
  {Thériault}}, \bibinfo {author} {\bibfnamefont {A.}~\bibnamefont
  {Gourgout}}, \bibinfo {author} {\bibfnamefont {M.-E.}\ \bibnamefont
  {Boulanger}}, \bibinfo {author} {\bibfnamefont {E.}~\bibnamefont
  {Lefrançois}}, \bibinfo {author} {\bibfnamefont {A.}~\bibnamefont {Ataei}},
  \bibinfo {author} {\bibfnamefont {F.}~\bibnamefont {Laliberté}}, \bibinfo
  {author} {\bibfnamefont {M.}~\bibnamefont {Dion}}, \bibinfo {author}
  {\bibfnamefont {J.-S.}\ \bibnamefont {Zhou}}, \bibinfo {author}
  {\bibfnamefont {S.}~\bibnamefont {Pyon}}, \bibinfo {author} {\bibfnamefont
  {T.}~\bibnamefont {Takayama}}, \bibinfo {author} {\bibfnamefont
  {H.}~\bibnamefont {Takagi}}, \bibinfo {author} {\bibfnamefont
  {N.}~\bibnamefont {Doiron-Leyraud}},\ and\ \bibinfo {author} {\bibfnamefont
  {L.}~\bibnamefont {Taillefer}},\ }\bibfield  {title} {\bibinfo {title}
  {Chiral phonons in the pseudogap phase of cuprates},\ }\href
  {https://doi.org/10.1038/s41567-020-0965-y} {\bibfield  {journal} {\bibinfo
  {journal} {Nat. Phys.}\ }\textbf {\bibinfo {volume} {16}},\ \bibinfo {pages}
  {1108} (\bibinfo {year} {2020})}\BibitemShut {NoStop}%
\bibitem [{\citenamefont {Chen}\ \emph {et~al.}(2020)\citenamefont {Chen},
  \citenamefont {Kivelson},\ and\ \citenamefont {Sun}}]{Chen2020}%
  \BibitemOpen
  \bibfield  {author} {\bibinfo {author} {\bibfnamefont {J.-Y.}\ \bibnamefont
  {Chen}}, \bibinfo {author} {\bibfnamefont {S.~A.}\ \bibnamefont {Kivelson}},\
  and\ \bibinfo {author} {\bibfnamefont {X.-Q.}\ \bibnamefont {Sun}},\
  }\bibfield  {title} {\bibinfo {title} {Enhanced thermal hall effect in nearly
  ferroelectric insulators},\ }\href
  {https://doi.org/10.1103/physrevlett.124.167601} {\bibfield  {journal}
  {\bibinfo  {journal} {Phys. Rev. Lett.}\ }\textbf {\bibinfo {volume} {124}},\
  \bibinfo {pages} {167601} (\bibinfo {year} {2020})}\BibitemShut {NoStop}%
\bibitem [{\citenamefont {Wang}\ \emph {et~al.}(2025)\citenamefont {Wang},
  \citenamefont {Xu}, \citenamefont {Liu}, \citenamefont {Wang}, \citenamefont
  {Zhang},\ and\ \citenamefont {Zhang}}]{WangAlteraxial2026}%
  \BibitemOpen
  \bibfield  {author} {\bibinfo {author} {\bibfnamefont {F.}~\bibnamefont
  {Wang}}, \bibinfo {author} {\bibfnamefont {J.}~\bibnamefont {Xu}}, \bibinfo
  {author} {\bibfnamefont {X.}~\bibnamefont {Liu}}, \bibinfo {author}
  {\bibfnamefont {H.}~\bibnamefont {Wang}}, \bibinfo {author} {\bibfnamefont
  {L.}~\bibnamefont {Zhang}},\ and\ \bibinfo {author} {\bibfnamefont
  {H.}~\bibnamefont {Zhang}},\ }\href {https://arxiv.org/abs/2512.07518}
  {\bibinfo {title} {Alteraxial phonons in collinear magnets}} (\bibinfo {year}
  {2025}),\ \Eprint {https://arxiv.org/abs/2512.07518} {arXiv:2512.07518
  [cond-mat.mtrl-sci]} \BibitemShut {NoStop}%
\bibitem [{\citenamefont {Coh}(2023)}]{Coh2023}%
  \BibitemOpen
  \bibfield  {author} {\bibinfo {author} {\bibfnamefont {S.}~\bibnamefont
  {Coh}},\ }\bibfield  {title} {\bibinfo {title} {Classification of materials
  with phonon angular momentum and microscopic origin of angular momentum},\
  }\href {https://doi.org/10.1103/PhysRevB.108.134307} {\bibfield  {journal}
  {\bibinfo  {journal} {Phys. Rev. B}\ }\textbf {\bibinfo {volume} {108}},\
  \bibinfo {pages} {134307} (\bibinfo {year} {2023})}\BibitemShut {NoStop}%
\bibitem [{\citenamefont {Steward}\ \emph {et~al.}(2023)\citenamefont
  {Steward}, \citenamefont {Fernandes},\ and\ \citenamefont
  {Schmalian}}]{Steward2023}%
  \BibitemOpen
  \bibfield  {author} {\bibinfo {author} {\bibfnamefont {C.~R.~W.}\
  \bibnamefont {Steward}}, \bibinfo {author} {\bibfnamefont {R.~M.}\
  \bibnamefont {Fernandes}},\ and\ \bibinfo {author} {\bibfnamefont
  {J.}~\bibnamefont {Schmalian}},\ }\bibfield  {title} {\bibinfo {title}
  {Dynamic paramagnon-polarons in altermagnets},\ }\href
  {https://doi.org/10.1103/PhysRevB.108.144418} {\bibfield  {journal} {\bibinfo
   {journal} {Phys. Rev. B}\ }\textbf {\bibinfo {volume} {108}},\ \bibinfo
  {pages} {144418} (\bibinfo {year} {2023})}\BibitemShut {NoStop}%
\bibitem [{\citenamefont {Bendin}\ \emph {et~al.}(2025)\citenamefont {Bendin},
  \citenamefont {Mook}, \citenamefont {Mertig},\ and\ \citenamefont
  {Neumann}}]{Bendin2025}%
  \BibitemOpen
  \bibfield  {author} {\bibinfo {author} {\bibfnamefont {H.}~\bibnamefont
  {Bendin}}, \bibinfo {author} {\bibfnamefont {A.}~\bibnamefont {Mook}},
  \bibinfo {author} {\bibfnamefont {I.}~\bibnamefont {Mertig}},\ and\ \bibinfo
  {author} {\bibfnamefont {R.~R.}\ \bibnamefont {Neumann}},\ }\href
  {10.48550/arXiv.2511.08357} {\bibinfo {title} {$d$-wave phonon angular
  momentum texture in altermagnets by magnon--phonon hybridization}} (\bibinfo
  {year} {2025}),\ \Eprint {https://arxiv.org/abs/2511.08357} {arXiv:2511.08357
  [cond-mat.mes-hall]} \BibitemShut {NoStop}%
\bibitem [{\citenamefont {Giustino}(2017)}]{Giustino2017}%
  \BibitemOpen
  \bibfield  {author} {\bibinfo {author} {\bibfnamefont {F.}~\bibnamefont
  {Giustino}},\ }\bibfield  {title} {\bibinfo {title} {Electron-phonon
  interactions from first principles},\ }\href
  {https://doi.org/10.1103/revmodphys.89.015003} {\bibfield  {journal}
  {\bibinfo  {journal} {Rev. Mod. Phys.}\ }\textbf {\bibinfo {volume} {89}},\
  \bibinfo {pages} {015003} (\bibinfo {year} {2017})}\BibitemShut {NoStop}%
\bibitem [{\citenamefont {Gerber}\ \emph {et~al.}(2017)\citenamefont {Gerber},
  \citenamefont {Yang}, \citenamefont {Zhu}, \citenamefont {Soifer},
  \citenamefont {Sobota}, \citenamefont {Rebec}, \citenamefont {Lee},
  \citenamefont {Jia}, \citenamefont {Moritz}, \citenamefont {Jia},
  \citenamefont {Gauthier}, \citenamefont {Li}, \citenamefont {Leuenberger},
  \citenamefont {Zhang}, \citenamefont {Chaix}, \citenamefont {Li},
  \citenamefont {Jang}, \citenamefont {Lee}, \citenamefont {Yi}, \citenamefont
  {Dakovski}, \citenamefont {Song}, \citenamefont {Glownia}, \citenamefont
  {Nelson}, \citenamefont {Kim}, \citenamefont {Chuang}, \citenamefont
  {Hussain}, \citenamefont {Moore}, \citenamefont {Devereaux}, \citenamefont
  {Lee}, \citenamefont {Kirchmann},\ and\ \citenamefont {Shen}}]{Gerber2017}%
  \BibitemOpen
  \bibfield  {author} {\bibinfo {author} {\bibfnamefont {S.}~\bibnamefont
  {Gerber}}, \bibinfo {author} {\bibfnamefont {S.-L.}\ \bibnamefont {Yang}},
  \bibinfo {author} {\bibfnamefont {D.}~\bibnamefont {Zhu}}, \bibinfo {author}
  {\bibfnamefont {H.}~\bibnamefont {Soifer}}, \bibinfo {author} {\bibfnamefont
  {J.~A.}\ \bibnamefont {Sobota}}, \bibinfo {author} {\bibfnamefont
  {S.}~\bibnamefont {Rebec}}, \bibinfo {author} {\bibfnamefont {J.~J.}\
  \bibnamefont {Lee}}, \bibinfo {author} {\bibfnamefont {T.}~\bibnamefont
  {Jia}}, \bibinfo {author} {\bibfnamefont {B.}~\bibnamefont {Moritz}},
  \bibinfo {author} {\bibfnamefont {C.}~\bibnamefont {Jia}}, \bibinfo {author}
  {\bibfnamefont {A.}~\bibnamefont {Gauthier}}, \bibinfo {author}
  {\bibfnamefont {Y.}~\bibnamefont {Li}}, \bibinfo {author} {\bibfnamefont
  {D.}~\bibnamefont {Leuenberger}}, \bibinfo {author} {\bibfnamefont
  {Y.}~\bibnamefont {Zhang}}, \bibinfo {author} {\bibfnamefont
  {L.}~\bibnamefont {Chaix}}, \bibinfo {author} {\bibfnamefont
  {W.}~\bibnamefont {Li}}, \bibinfo {author} {\bibfnamefont {H.}~\bibnamefont
  {Jang}}, \bibinfo {author} {\bibfnamefont {J.-S.}\ \bibnamefont {Lee}},
  \bibinfo {author} {\bibfnamefont {M.}~\bibnamefont {Yi}}, \bibinfo {author}
  {\bibfnamefont {G.~L.}\ \bibnamefont {Dakovski}}, \bibinfo {author}
  {\bibfnamefont {S.}~\bibnamefont {Song}}, \bibinfo {author} {\bibfnamefont
  {J.~M.}\ \bibnamefont {Glownia}}, \bibinfo {author} {\bibfnamefont
  {S.}~\bibnamefont {Nelson}}, \bibinfo {author} {\bibfnamefont {K.~W.}\
  \bibnamefont {Kim}}, \bibinfo {author} {\bibfnamefont {Y.-D.}\ \bibnamefont
  {Chuang}}, \bibinfo {author} {\bibfnamefont {Z.}~\bibnamefont {Hussain}},
  \bibinfo {author} {\bibfnamefont {R.~G.}\ \bibnamefont {Moore}}, \bibinfo
  {author} {\bibfnamefont {T.~P.}\ \bibnamefont {Devereaux}}, \bibinfo {author}
  {\bibfnamefont {W.-S.}\ \bibnamefont {Lee}}, \bibinfo {author} {\bibfnamefont
  {P.~S.}\ \bibnamefont {Kirchmann}},\ and\ \bibinfo {author} {\bibfnamefont
  {Z.-X.}\ \bibnamefont {Shen}},\ }\bibfield  {title} {\bibinfo {title}
  {Femtosecond electron-phonon lock-in by photoemission and x-ray free-electron
  laser},\ }\href {https://doi.org/10.1126/science.aak9946} {\bibfield
  {journal} {\bibinfo  {journal} {Science}\ }\textbf {\bibinfo {volume}
  {357}},\ \bibinfo {pages} {71} (\bibinfo {year} {2017})}\BibitemShut
  {NoStop}%
\bibitem [{\citenamefont {Valla}\ \emph {et~al.}(1999)\citenamefont {Valla},
  \citenamefont {Fedorov}, \citenamefont {Johnson},\ and\ \citenamefont
  {Hulbert}}]{Valla1999}%
  \BibitemOpen
  \bibfield  {author} {\bibinfo {author} {\bibfnamefont {T.}~\bibnamefont
  {Valla}}, \bibinfo {author} {\bibfnamefont {A.~V.}\ \bibnamefont {Fedorov}},
  \bibinfo {author} {\bibfnamefont {P.~D.}\ \bibnamefont {Johnson}},\ and\
  \bibinfo {author} {\bibfnamefont {S.~L.}\ \bibnamefont {Hulbert}},\
  }\bibfield  {title} {\bibinfo {title} {Many-body effects in angle-resolved
  photoemission: Quasiparticle energy and lifetime of a mo(110) surface
  state},\ }\href {https://doi.org/10.1103/physrevlett.83.2085} {\bibfield
  {journal} {\bibinfo  {journal} {Phys. Rev. Lett.}\ }\textbf {\bibinfo
  {volume} {83}},\ \bibinfo {pages} {2085} (\bibinfo {year}
  {1999})}\BibitemShut {NoStop}%
\bibitem [{\citenamefont {Lanzara}\ \emph {et~al.}(2001)\citenamefont
  {Lanzara}, \citenamefont {Bogdanov}, \citenamefont {Zhou}, \citenamefont
  {Kellar}, \citenamefont {Feng}, \citenamefont {Lu}, \citenamefont {Yoshida},
  \citenamefont {Eisaki}, \citenamefont {Fujimori}, \citenamefont {Kishio},
  \citenamefont {Shimoyama}, \citenamefont {Noda}, \citenamefont {Uchida},
  \citenamefont {Hussain},\ and\ \citenamefont {Shen}}]{Lanzara2001}%
  \BibitemOpen
  \bibfield  {author} {\bibinfo {author} {\bibfnamefont {A.}~\bibnamefont
  {Lanzara}}, \bibinfo {author} {\bibfnamefont {P.~V.}\ \bibnamefont
  {Bogdanov}}, \bibinfo {author} {\bibfnamefont {X.~J.}\ \bibnamefont {Zhou}},
  \bibinfo {author} {\bibfnamefont {S.~A.}\ \bibnamefont {Kellar}}, \bibinfo
  {author} {\bibfnamefont {D.~L.}\ \bibnamefont {Feng}}, \bibinfo {author}
  {\bibfnamefont {E.~D.}\ \bibnamefont {Lu}}, \bibinfo {author} {\bibfnamefont
  {T.}~\bibnamefont {Yoshida}}, \bibinfo {author} {\bibfnamefont
  {H.}~\bibnamefont {Eisaki}}, \bibinfo {author} {\bibfnamefont
  {A.}~\bibnamefont {Fujimori}}, \bibinfo {author} {\bibfnamefont
  {K.}~\bibnamefont {Kishio}}, \bibinfo {author} {\bibfnamefont {J.-I.}\
  \bibnamefont {Shimoyama}}, \bibinfo {author} {\bibfnamefont {T.}~\bibnamefont
  {Noda}}, \bibinfo {author} {\bibfnamefont {S.}~\bibnamefont {Uchida}},
  \bibinfo {author} {\bibfnamefont {Z.}~\bibnamefont {Hussain}},\ and\ \bibinfo
  {author} {\bibfnamefont {Z.-X.}\ \bibnamefont {Shen}},\ }\bibfield  {title}
  {\bibinfo {title} {Evidence for ubiquitous strong electron–phonon coupling
  in high-temperature superconductors},\ }\href
  {https://doi.org/10.1038/35087518} {\bibfield  {journal} {\bibinfo  {journal}
  {Nature}\ }\textbf {\bibinfo {volume} {412}},\ \bibinfo {pages} {510}
  (\bibinfo {year} {2001})}\BibitemShut {NoStop}%
\bibitem [{\citenamefont {Först}\ \emph {et~al.}(2011)\citenamefont {Först},
  \citenamefont {Manzoni}, \citenamefont {Kaiser}, \citenamefont {Tomioka},
  \citenamefont {Tokura}, \citenamefont {Merlin},\ and\ \citenamefont
  {Cavalleri}}]{Foerst2011}%
  \BibitemOpen
  \bibfield  {author} {\bibinfo {author} {\bibfnamefont {M.}~\bibnamefont
  {Först}}, \bibinfo {author} {\bibfnamefont {C.}~\bibnamefont {Manzoni}},
  \bibinfo {author} {\bibfnamefont {S.}~\bibnamefont {Kaiser}}, \bibinfo
  {author} {\bibfnamefont {Y.}~\bibnamefont {Tomioka}}, \bibinfo {author}
  {\bibfnamefont {Y.}~\bibnamefont {Tokura}}, \bibinfo {author} {\bibfnamefont
  {R.}~\bibnamefont {Merlin}},\ and\ \bibinfo {author} {\bibfnamefont
  {A.}~\bibnamefont {Cavalleri}},\ }\bibfield  {title} {\bibinfo {title}
  {Nonlinear phononics as an ultrafast route to lattice control},\ }\href
  {https://doi.org/10.1038/nphys2055} {\bibfield  {journal} {\bibinfo
  {journal} {Nat. Phys.}\ }\textbf {\bibinfo {volume} {7}},\ \bibinfo {pages}
  {854} (\bibinfo {year} {2011})}\BibitemShut {NoStop}%
\bibitem [{\citenamefont {Mankowsky}\ \emph {et~al.}(2016)\citenamefont
  {Mankowsky}, \citenamefont {Först},\ and\ \citenamefont
  {Cavalleri}}]{Mankowsky2016}%
  \BibitemOpen
  \bibfield  {author} {\bibinfo {author} {\bibfnamefont {R.}~\bibnamefont
  {Mankowsky}}, \bibinfo {author} {\bibfnamefont {M.}~\bibnamefont {Först}},\
  and\ \bibinfo {author} {\bibfnamefont {A.}~\bibnamefont {Cavalleri}},\
  }\bibfield  {title} {\bibinfo {title} {Non-equilibrium control of complex
  solids by nonlinear phononics},\ }\href
  {https://doi.org/10.1088/0034-4885/79/6/064503} {\bibfield  {journal}
  {\bibinfo  {journal} {Rep. Progr. Phys.}\ }\textbf {\bibinfo {volume} {79}},\
  \bibinfo {pages} {064503} (\bibinfo {year} {2016})}\BibitemShut {NoStop}%
\bibitem [{\citenamefont {Li}\ \emph {et~al.}(2012)\citenamefont {Li},
  \citenamefont {Ren}, \citenamefont {Wang}, \citenamefont {Zhang},
  \citenamefont {Hänggi},\ and\ \citenamefont {Li}}]{Li2012}%
  \BibitemOpen
  \bibfield  {author} {\bibinfo {author} {\bibfnamefont {N.}~\bibnamefont
  {Li}}, \bibinfo {author} {\bibfnamefont {J.}~\bibnamefont {Ren}}, \bibinfo
  {author} {\bibfnamefont {L.}~\bibnamefont {Wang}}, \bibinfo {author}
  {\bibfnamefont {G.}~\bibnamefont {Zhang}}, \bibinfo {author} {\bibfnamefont
  {P.}~\bibnamefont {Hänggi}},\ and\ \bibinfo {author} {\bibfnamefont
  {B.}~\bibnamefont {Li}},\ }\bibfield  {title} {\bibinfo {title} {Colloquium:
  Phononics: Manipulating heat flow with electronic analogs and beyond},\
  }\href {https://doi.org/10.1103/revmodphys.84.1045} {\bibfield  {journal}
  {\bibinfo  {journal} {Rev. Mod. Phys.}\ }\textbf {\bibinfo {volume} {84}},\
  \bibinfo {pages} {1045} (\bibinfo {year} {2012})}\BibitemShut {NoStop}%
\bibitem [{\citenamefont {Wehmeyer}\ \emph {et~al.}(2017)\citenamefont
  {Wehmeyer}, \citenamefont {Yabuki}, \citenamefont {Monachon}, \citenamefont
  {Wu},\ and\ \citenamefont {Dames}}]{Wehmeyer2017}%
  \BibitemOpen
  \bibfield  {author} {\bibinfo {author} {\bibfnamefont {G.}~\bibnamefont
  {Wehmeyer}}, \bibinfo {author} {\bibfnamefont {T.}~\bibnamefont {Yabuki}},
  \bibinfo {author} {\bibfnamefont {C.}~\bibnamefont {Monachon}}, \bibinfo
  {author} {\bibfnamefont {J.}~\bibnamefont {Wu}},\ and\ \bibinfo {author}
  {\bibfnamefont {C.}~\bibnamefont {Dames}},\ }\bibfield  {title} {\bibinfo
  {title} {Thermal diodes, regulators, and switches: Physical mechanisms and
  potential applications},\ }\href {https://doi.org/10.1063/1.5001072}
  {\bibfield  {journal} {\bibinfo  {journal} {Appl. Phys. Rev.}\ }\textbf
  {\bibinfo {volume} {4}},\ \bibinfo {pages} {041304} (\bibinfo {year}
  {2017})}\BibitemShut {NoStop}%
\bibitem [{\citenamefont {Miao}\ \emph {et~al.}(2018)\citenamefont {Miao},
  \citenamefont {Zhang}, \citenamefont {Wang}, \citenamefont {Meyers},
  \citenamefont {Said}, \citenamefont {Wang}, \citenamefont {Shi},
  \citenamefont {Weng}, \citenamefont {Fang},\ and\ \citenamefont
  {Dean}}]{Miao2018}%
  \BibitemOpen
  \bibfield  {author} {\bibinfo {author} {\bibfnamefont {H.}~\bibnamefont
  {Miao}}, \bibinfo {author} {\bibfnamefont {T.}~\bibnamefont {Zhang}},
  \bibinfo {author} {\bibfnamefont {L.}~\bibnamefont {Wang}}, \bibinfo {author}
  {\bibfnamefont {D.}~\bibnamefont {Meyers}}, \bibinfo {author} {\bibfnamefont
  {A.}~\bibnamefont {Said}}, \bibinfo {author} {\bibfnamefont {Y.}~\bibnamefont
  {Wang}}, \bibinfo {author} {\bibfnamefont {Y.}~\bibnamefont {Shi}}, \bibinfo
  {author} {\bibfnamefont {H.}~\bibnamefont {Weng}}, \bibinfo {author}
  {\bibfnamefont {Z.}~\bibnamefont {Fang}},\ and\ \bibinfo {author}
  {\bibfnamefont {M.}~\bibnamefont {Dean}},\ }\bibfield  {title} {\bibinfo
  {title} {Observation of double weyl phonons in parity-breaking fesi},\ }\href
  {https://doi.org/10.1103/physrevlett.121.035302} {\bibfield  {journal}
  {\bibinfo  {journal} {Phys. Rev. Lett.}\ }\textbf {\bibinfo {volume} {121}},\
  \bibinfo {pages} {035302} (\bibinfo {year} {2018})}\BibitemShut {NoStop}%
\bibitem [{\citenamefont {Baron}(2020)}]{baron2020}%
  \BibitemOpen
  \bibfield  {author} {\bibinfo {author} {\bibfnamefont {A.~Q.~R.}\
  \bibnamefont {Baron}},\ }\href {https://arxiv.org/abs/1504.01098} {\bibinfo
  {title} {Introduction to high-resolution inelastic x-ray scattering}}
  (\bibinfo {year} {2020}),\ \Eprint {https://arxiv.org/abs/1504.01098}
  {arXiv:1504.01098 [cond-mat.mtrl-sci]} \BibitemShut {NoStop}%
\bibitem [{\citenamefont {Senga}\ \emph {et~al.}(2019)\citenamefont {Senga},
  \citenamefont {Suenaga}, \citenamefont {Barone}, \citenamefont {Morishita},
  \citenamefont {Mauri},\ and\ \citenamefont {Pichler}}]{Senga2019}%
  \BibitemOpen
  \bibfield  {author} {\bibinfo {author} {\bibfnamefont {R.}~\bibnamefont
  {Senga}}, \bibinfo {author} {\bibfnamefont {K.}~\bibnamefont {Suenaga}},
  \bibinfo {author} {\bibfnamefont {P.}~\bibnamefont {Barone}}, \bibinfo
  {author} {\bibfnamefont {S.}~\bibnamefont {Morishita}}, \bibinfo {author}
  {\bibfnamefont {F.}~\bibnamefont {Mauri}},\ and\ \bibinfo {author}
  {\bibfnamefont {T.}~\bibnamefont {Pichler}},\ }\bibfield  {title} {\bibinfo
  {title} {Position and momentum mapping of vibrations in graphene
  nanostructures},\ }\href {https://doi.org/10.1038/s41586-019-1477-8}
  {\bibfield  {journal} {\bibinfo  {journal} {Nature}\ }\textbf {\bibinfo
  {volume} {573}},\ \bibinfo {pages} {247} (\bibinfo {year}
  {2019})}\BibitemShut {NoStop}%
\end{thebibliography}
\end{document}